\documentclass{aa}

\usepackage{graphicx}
\usepackage{txfonts}
\usepackage{natbib}
\usepackage{amssymb}
\usepackage{gensymb}

\begin{document}

   \title{High-resolution international LOFAR observations of 4C~43.15}
   \subtitle{Spectral ages and injection indices in a high-z radio galaxy}
   \titlerunning{High-resolution international LOFAR observations of 4C~43.15}
   \authorrunning{F. Sweijen et al.}

   \author{
F. Sweijen,$^{1}$\thanks{E-mail: sweijen@strw.leidenuniv.nl}
L. K. Morabito,$^{2}$
J. Harwood,$^3$
R. J. van Weeren,$^{1}$
H. J. A. R\"ottgering,$^{1}$,
J. R. Callingham,$^{1,4}$
N. Jackson,${^5}$
G. Miley,$^{^1}$
J. Moldon,$^{6,7}$}

   \institute{
$^{1}$Sterrewacht Leiden, University of Leiden, 2300 RA, Leiden, The Netherlands\\
$^{2}$Centre for Extragalactic Astronomy, Department of Physics, Durham University, DH1 3LE, UK\\
$^{3}$Centre for Astrophysics Research, School of Physics, Astronomy and Mathematics, University of Hertfordshire, College Lane\\
\ Hatfield, Hertfordshire AL10 9AB, UK\\
$^4$ASTRON, Netherlands Institute for Radio Astronomy, Oude Hoogeveensedijk 4, Dwingeloo NL-7991 PD, The Netherlands\\
$^{5}$University of Manchester, School of Physics and Astronomy, JodrellBank Centre for Astrophysics, Oxford Road, Manchester M13 9PL,UK\\
$^{6}$Instituto de Astrof\'isica de Andaluc\'ia (IAA, CSIC), Glorieta de las Astronom\'ia, s/n, E-18008 Granada, Spain\\
$^{7}$Jodrell Bank Centre for Astrophysics, School of Physics and Astronomy, University of Manchester, Manchester M13 9PL, UK
}

   \date{Received September 15, 1996; accepted March 16, 1997}

 
  \abstract{It has long been known that radio sources with the steepest spectra are preferentially associated with the most distant galaxies, the $\alpha-z$ relation, but the reason for this relation is an open question. The spatial distribution of spectra in high-z radio sources can be used to study this relation, and low-frequency observations are particularly important in understanding the particle acceleration and injection mechanisms. However, the small angular sizes of high-z sources together with the inherently low resolution of low-frequency radio telescopes until now has prevented high angular resolution low-frequency observations of distant objects. Here we present subarcsecond observations of a $z = 2.4$ radio galaxy at frequencies between $121$ MHz and $166$ MHz. We measure the spatial distribution of spectra, and discuss the implications for models of the $\alpha-z$ relation. We targeted 4C 43.15 with the High Band Antennas (HBAs) of the \textit{International LOFAR Telescope} (ILT) with a range of baselines up to $1300\ \mathrm{km}$. At the central frequency of $143$ MHz we achieve an angular resolution of $\sim 0.3''$. By complementing our data with archival \textit{Very Large Array} (VLA) data we study the spectral index distribution across 4C 43.15 between $55\ \mathrm{MHz}$ and $8.4\ \mathrm{GHz}$ at resolutions of $0.4''$ and $0.9''$. With a magnetic field strength of $B = 5.2$ nT and fitted injection indices of $\alpha^\mathrm{north}_\mathrm{inj} = -0.8$ and $\alpha^\mathrm{south}_\mathrm{inj} = -0.6$, fitting a Tribble spectral ageing model results in a spectral age of $\tau_\mathrm{spec} = 1.1 \pm 0.1$ Myr. We conclude that our data on 4C 43.15 indicates that inverse Compton losses could become comparable to or exceed synchrotron losses at higher redshifts and that inverse Compton losses could be a viable explanation for the $\alpha-z$ relation. Statistical studies of these objects will become possible in the future when wide-area subarcsecond surveys start.}

   \keywords{galaxies: active -- galaxies: evolution -- galaxies: high-redshift -- radio continuum: galaxies -- radiation mechanisms: non-thermal}

   \maketitle
%

\section{Introduction}
High-redshift radio galaxies are important probes of galaxy evolution as they allow the study of the (co)evolution of black holes and their host galaxies over cosmic time. A correlation that has frequently been used to search for high-redshift radio galaxies, but is still under debate, is the $\alpha-z$ relation, where  radio sources with steeper spectra appear to be preferentially associated with higher redshift sources \citep{Tielens1979,DeBreuck2000,Miley2008}. One of the mechanisms that has been suggested to explain this relation is that this is the result of the increased inverse Compton (IC) losses at higher redshifts, due to the higher energy density of the cosmic microwave background (CMB) (e.g. \citealt{Ghisellini2014} or \citealt{Klamer2006}). Recently, \cite{Morabito2018} found that IC losses could indeed be the primary driver behind the $\alpha-z$ relation, by modelling a large number of sources over a wide range of redshifts using the \textit{Broadband Radio Astronomy ToolS} (\textsc{brats}) software package\footnote{\url{http://www.askanastronomer.co.uk/brats/}} \citep{Harwood2013, Harwood2015}. There are two other scenarios that could result in this relation,  one connected with observational biases and the other  related to the environment of the galaxy. \cite{Blundell1999} studied the correlations between  radio luminosity, linear size, spectral index, and redshift for flux-density--limited surveys, and propose that the observed $\alpha-z$ relation follows from a $P-\alpha$ relation, where $P$ is the radio power. The Malmquist bias will then naturally drive an $\alpha-z$ correlation as the most powerful sources are the easiest to detect. \cite{Klamer2006} suggest that an increased ambient density around radio sources in the early Universe could be an important factor in steepening the spectrum. A denser medium at high redshift would keep the radio lobes more confined, and would result in a higher surface brightness for older radio emission coming from the low-energy electron populations. If at higher redshifts radio galaxies   preferentially lie in denser environments, these effects could lead to the observed relation. Another effect from an increased ambient density is related to the acceleration of electrons in or near the hotspot. Based on the findings by  \cite{Kirk1987}, among others,  that the spectral index is sensitive to properties of the shock front near the hotspot, \cite{Athreya1998} proposed that a radio jet working against higher ambient densities would already have a  steeper initial spectrum of electrons being produced before flowing back into the lobes and steepening more due to further energy losses. \cite{Gopal-Krishna2012} argue against steeper injection indices based on the absence of a correlation between spectral index and rotation measure (which probes the density) in a sample of compact steep spectrum (CSS) sources from \cite{Murgia2002} which have a median injection index of $\alpha=-0.63$.

Studying the spatial distribution of spectra over as wide a frequency range as possible is crucial in disentangling the various processes that produce the radio emission, such as synchrotron ageing, IC losses, and evolution. The rate at which energy is lost depends on the energy of the radiating or scattering electrons. These energy losses over time cause the spectrum to deviate from a power law and introduce curvature that becomes more pronounced at higher frequencies. This curving of the spectrum is referred to as the spectral ageing of the source. It is for this reason that measurements over a large frequency bandwidth are required to properly characterise the spectrum. For decades work has been done using measurements across the radio spectrum to study the evolution of radio galaxies \citep[e.g.][]{MyersSpangler1985,AlexanderLeahy1987,Carilli1991,Jamrozy2008,Harwood2013, Harwood2016, Mahatma2019}. With multiple measurements it is possible  to fit a spectral ageing model to the spectrum to obtain an estimate of the spectral age of the plasma. Low-frequency data plays a crucial role in such modelling because it constrains the injection index $\alpha_\mathrm{inj}$, which is  the spectral index of the young electrons in hotspots, recently (re)accelerated through the interaction of the jet with the surrounding medium. This part of the spectrum will be closer to the original power law, while at higher frequencies the spectrum will steepen due to ageing of the electrons.

To fully understand the evolution of these sources, spatially resolved studies spanning the entire radio spectrum over a range of redshifts are required. Until recently, high angular resolution observations of these high-redshift objects have been limited to gigahertz frequencies or higher, using facilities such as the Very Large Array (VLA), or the technique of very long baseline interferometry (VLBI). Large nearby sources such as Cygnus A are easily resolved at most frequencies (see e.g. \citealt{McKean2016} or \citealt{Perley1984} in the case of Cygnus A at megahertz and gigahertz frequencies, respectively). However, the typically poor angular resolution of low-frequency radio telescopes prevents such detailed views of their higher-redshift brethren, and has therefore limited spectral studies of distant AGNs to numerical models \citep[e.g.][]{Huarte-Espinosa2011, Cielo2018}. \cite{Harwood2017a} shows that modelling sources using only their integrated spectrum is unreliable and cannot recover the correct value for properties such as the injection index and robust spectral ages, highlighting the need for high angular resolution observations. The LOw-Frequency ARray (LOFAR; \citealt{Haarlem2013}) is a low-frequency interferometer. It can observe between $10$ and $90\ \mathrm{MHz}$ with the Low Band Antennas (LBAs) and between $110$ and $240\ \mathrm{MHz}$ with the High Band Antennas (HBAs). With its international stations it is possible to venture into subarcsecond territory at frequencies below $ 240\ \mathrm{MHz}$. This opens up the window for (highly) spatially resolved studies of objects at low frequencies  \citep[e.g.][]{Morabito2016, Varenius2014, Varenius2016, Ramirez-Olivencia2018, Harris2019}.

This paper is a follow-up study at $143\ \mathrm{MHz}$ of the $55\ \mathrm{MHz}$ study presented in \cite{Morabito2016}. 4C 43.15 is part of a small sample of high-redshift radio galaxies to be studied in detail at low frequencies. These sources were selected to investigate the $\alpha-z$ relation because of their (apparent) ultra-steep spectra. 4C 43.15 is a high-redshift radio galaxy (HzRG) with a redshift of $z = 2.429$ \citep{McCarthy1991}. It has an integrated spectral index between $365\ \mathrm{MHz}$ and $1.4\ \mathrm{GHz}$ of $\alpha = -1.1$. \cite{Morabito2016} showed that for 4C 43.15 this ultra-steep spectral index is the result of a break in the spectrum at intermediate frequencies and steepening at higher frequencies. Here we add $143\ \mathrm{MHz}$ LOFAR data and, combined with archival VLA data, present a spatially resolved spectral study of a high-redshift galaxy, from $55\ \mathrm{MHz}$ to $8.46\ \mathrm{GHz}$. We use the LOFAR LBA data presented by \cite{Morabito2016}.

In this paper we try to assess whether IC losses can be a main contributor to the steepness of the spectrum of 4C 43.15, using high spatial resolution low-frequency observations. Section 2 describes the observations that were used. In Section 3 the data reduction and spectral modelling are described. Section 4 describes the results that were obtained. Finally, sections 5 and 6 discuss the results in context of the $\alpha-z$ relation and present the conclusions.

The assumed cosmology is that of the 2015 Planck results \citep{Ade2016}, with $H_0 = 67.8\ \mathrm{km}\mathrm{s}^{-1}\mathrm{Mpc}^{-1}$, $\Omega_\mathrm{m} = 0.308$, and $\Omega_\Lambda = 0.692$. At the redshift of 4C 43.15 $1''$ corresponds to $8.3\ \mathrm{kpc}$. Furthermore, the spectral index $\alpha$ is defined as $S_\nu \propto \nu^{\alpha}$ and $S_\nu$ is the flux density at a given frequency $\nu$.

\section{Observations}
The LOFAR HBA observations of 4C 43.15 presented in this paper were carried out on 21 January 2016, starting at $19$:$16$ UT (PI: Morabito, LT5\_006, L427100). The phase centre of the beam was pointed at J2000 $\alpha=7$h$35$m$22.44$s, $\delta=43\degree44'24.72''$, and the target was observed with a standard eight-hour pointing. The HBA\_DUAL\_INNER configuration was used, including international stations. In total 47 stations participated in the observation:  24 core stations and  14 remote stations, hereafter referred to  as the Dutch array, and 9 international stations in Germany, France, Sweden, and the UK. The international stations were DE601, DE602, DE603, DE604, DE605, DE609, FR606, SE607, and UK608. This provides a longest projected baseline of $1292\ \mathrm{km}$, giving a synthesised beam size of the order of $\sim 0.3''$ at $151\ \mathrm{MHz}$. Data were recorded at $1\ \mathrm{s}$, $3.051\ \mathrm{kHz}$ resolution and averaged to $2\ \mathrm{s}$, $12.207\ \mathrm{kHz}$ resolution before being uploaded to the long-term archive (LTA). A total bandwidth of $47.1\ \mathrm{MHz}$ covered the frequency range from $120.3\ \mathrm{MHz}$ to $167.4\ \mathrm{MHz}$.

Observations of two compact flux density calibrators, 3C\,147 and 3C\,48, bookended the target observation of 4C 43.15. The calibrator scans were $15$ minutes long and had the same configuration as the target observations.

Archival VLA observations in C band ($4535.1$ MHz and $4885.1$ MHz) and X band ($8414.9$ MHz) were used: project codes AC0374 \citep{Carilli1997} and AK0410, respectively. Each spectral window had $50$ MHz of bandwidth. Both projects had data taken with the VLA in A configuration. Observations were carried out on 19 March  1994 (AC0374) and 31 August  1995 (AK0410). The total on-source time was $28$ minutes for AC0374 and $40$ minutes for AK0410.

\section{Data reduction}
The LOFAR HBA raw data (L427100) were fully reduced using the LOFAR data reduction software, and the VLA data (projects AC0374 and AK0410) were retrieved from the archive and reduced again with the \textit{Common Astronomy Software Applications} (CASA) package. The LBA data were already reduced and we used the radio map obtained by \citet{Morabito2016}. This image is shown in Fig.~\ref{fig:lbamap}.

Data reduction of the HBA data consisted of three stages. First, the shorter baselines from the Dutch array were calibrated using the prefactor pipeline\footnote{\url{https://github.com/lofar-astron/prefactor}}. Subsequently, a beta version of the long baseline pipeline\footnote{\url{https://github.com/lmorabit/long_baseline_pipeline}} was used to calibrate the international stations. Finally, self-calibration was performed on the complete data set. The prefactor pipeline was run with version 2.19 of the LOFAR software.

The prefactor pipeline consists of two parts: the calibrator  and the target. First the calibrators were processed using the calibrator part of the pipeline, solving for a diagonal gain and the common rotation angle. This was followed by clock--TEC separation, which separates the phase effects due to clock drifts from those due to the ionosphere (TEC) (see e.g. \citealt{Weeren2016}). Additionally, the phase offset between X and Y polarisations was determined. Both the 3C 48 and 3C 147 scan were processed. The phase solutions for 3C 48 showed inexplicable shifts in time, which 3C 147 did not. Hence the solutions obtained from 3C 147 were used. It is important to note that the calibrator pipeline derived solutions for {all the stations} (i.e. Dutch {and} international stations).

\subsection{Dutch LOFAR calibration}
Following the calibrator reduction, the target field was processed with the target part of the pipeline. This first corrects the clock offsets, and applies gain and primary beam solutions. Finally, a correction for the rotation angle was determined and applied. The data were then averaged to $4\ \mathrm{s}$ time resolution and $48.82\ \mathrm{kHz}$ frequency resolution (four channels per sub-band), before a phase-only calibration on the target was carried out using a sky model from the Tata Institute of Fundamental reasearch (TIFR) \textit{Giant Metre Wave Radio Telescope} (GMRT) Sky Survey (TGSS) \citep{Intema2017} for the target field. This time and frequency resolution is a balance between keeping data size down and retaining the ability to apply or find corrections with the required precision. From the TGSS sky model, phase solutions for only the core and remote stations are obtained.

\subsection{International LOFAR calibration}
The LOFAR international stations were calibrated using a beta version of the long-baseline pipeline being developed by Morabito et al. (in prep.), using version $2.20$ of the LOFAR software. In this section we  describe the calibration strategy used in this version of the pipeline and the subsequent self-calibration that was carried out by hand. The starting point for the pipeline is the gain solutions from the calibrator part of prefactor, the phase solutions from the target part, and a user-supplied list of potential in-field calibrators selected from the Long Baseline Calibrator Survey (LBCS, \cite{Moldon2015, Jackson2016}). We started with the raw uncalibrated data as input, with its original temporal and spectral resolution.

\subsubsection{Apply prefactor solutions}
As a first step the input data were matched against the available solutions produced by calibrating the Dutch array. This ensured that the best available solutions were applied to each sub-band, and that sub-bands for which there were no matching solutions were excluded from further processing. The matching sub-bands were then copied to the working directory. The international station DE603, $\sim 400\ \mathrm{km}$ from the core, was flagged because its signal was unusable. This resulted in a slightly diminished uv-coverage at the shorter international baselines.

Next the amplitude and clock solutions derived from the calibrator data were interpolated to each individual sub-band and stored in a separate calibration table. Solutions were derived for all stations.

Finally, the \textit{New Default Pre-Processing Pipeline} (NDPPP; \citealt{DPPP}) was used to apply all initial corrections to the data. At this stage the data had been corrected for station clock offsets, gain corrections, beam corrections, and phase corrections, in this order.

\subsubsection{In-field calibrator selection}
\begin{figure}
        \includegraphics[width=\columnwidth]{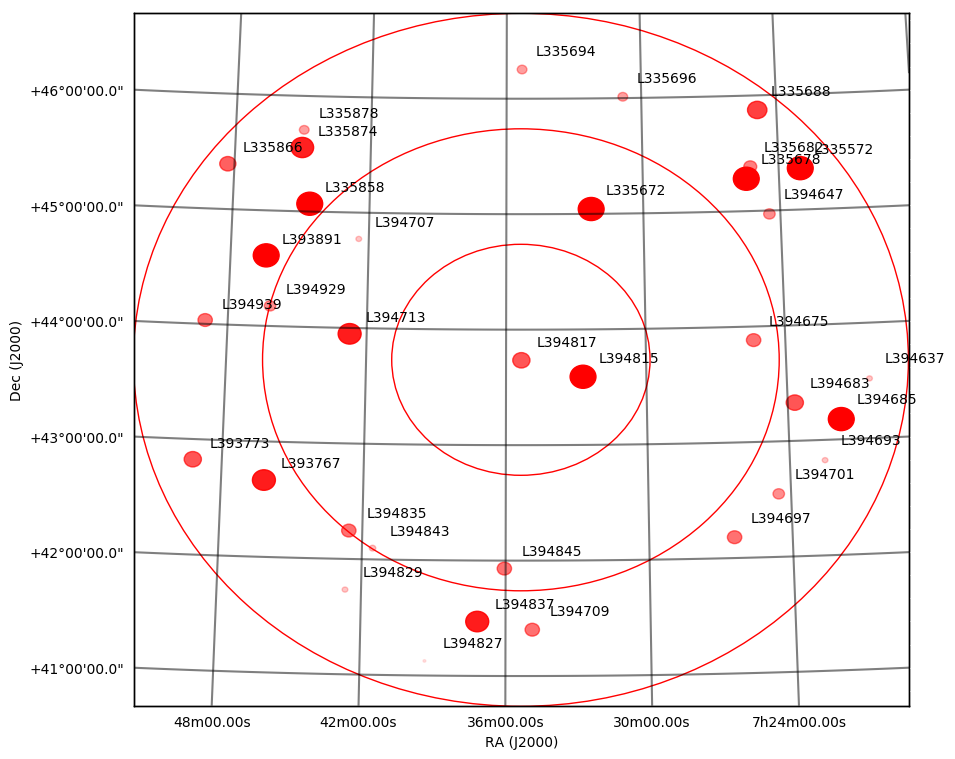}
        \caption{Overview of potential calibrators within $3\degree$ of the target's coordinates. The red circles are drawn at $1\degree$, $2\degree$, and $3\degree$ radii. Three of the closest calibrators were chosen to be used during calibration: L394813, L394815, and L394817. Bigger, redder dots mean more stations have coherent phases on the calibrator in question.}
    \label{fig:lbcs_plot}
\end{figure}

Before running this version of the pipeline, however, one or more potentially suitable calibrator(s) within the 4C 43.15 field needed to be determined. These calibrators were selected manually using a Python script that queries the LBCS database and presents potential calibrator sources, as shown in Fig.~\ref{fig:lbcs_plot}. Using this plot, the sources indicated with observation IDs L394813, L394815, and L394817 were selected to be used in the pipeline. This is only an initial selection. The pipeline will decide on the best calibrator by means of closure phases.

When the initial corrections were applied, a new data set for each selected calibrator was created by phase shifting to their respective positions. After the data were phase-shifted to each calibrator, data from all core stations were added coherently to create a new station ST001, which is temporarily equivalent to an additional more sensitive station, and finally the data were averaged to $8\ \mathrm{s}$ time resolution and one channel per sub-band ($195.312\ \mathrm{kHz}$) frequency resolution. ST001 was not used in the final imaging process because of its small field of view. The final data sets for each calibrator contain the full bandwidth and span the entire observation. By means of the closure phase, the relative quality of the calibrators to each other was assessed. The benefit of closure phases is that the sum of phases on a triangle of telescopes is independent of atmospheric errors \citep{Jennison1958}.

Calculated closure phases were then unwrapped and the mean value of the squared gradient was calculated. The best calibrator candidate was then selected as the source with the lowest mean value. For 4C43.15 the best in-field calibrator candidate turned out to be L394815. It is located at $\alpha = 7$h$32$m$43.72$s and $\delta = 43\degree35'40.50''$, about half a degree away from the phase centre. Stations DE601, DE605, and ST001 formed the triangle on which the closure phase was measured. The statistic as measured by the process described earlier was $0.23$. For L394813 and L394817 the values were $1.55$ and $0.42$, respectively.

\subsubsection{Phase reference calibration}
In the next step the in-field calibrator is calibrated against a model to obtain gain solutions. The initial model for the calibrator was a circular Gaussian point source, $0.1''$ in size. Solution intervals were chosen equal to the resolution of the data, $8\ \mathrm{s}$ and 1 channel or $195.312\ \mathrm{kHz}$ (i.e. a single time and frequency slot).

After this calibration the gains are applied to the target data, where all core stations are calibrated with the solutions for the phased-up station, ST001. Solutions were inspected by eye and found to smoothly vary in both time and frequency. The phase slope with respect to frequency is the delay (e.g. $306\ \mathrm{ns}$ on the remote station RS306 and $1.2\ \mu\mathrm{s}$ on the international station UK608). The following calibration solutions were then applied to the raw data, in the listed order:
\begin{enumerate}
        \item Phase offsets as a function of time and frequency, obtained from the clock offsets between the stations, determined by the prefactor calibrator pipeline;
    \item Amplitude solutions for all stations from the prefactor calibrator pipeline to set the global flux density scale. We use the flux scale by \cite{Scaife2012};
    \item Primary beam corrections to take into account the attenuating effect of the primary beam as a function of distance to the field phase centre;
    \item The common rotation angle found by RMextract\footnote{\url{https://github.com/lofar-astron/RMextract}} \citep{Mevius2018}, which corrects for Faraday rotation;
    \item Direction-independent phase solutions from the prefactor target pipeline, obtained by calibrating against a TGSS sky model;
    \item Gain solutions obtained from the in-field calibrator.
\end{enumerate}
Finally, the data were averaged to $16\ \mathrm{s}$ and $195.312\ \mathrm{kHz}$ (1 channel per sub-band), and were  compressed using Dysco \citep{Offringa2016} into $65$ GB of calibrated data compared to $7$ TB of raw data.

\subsection{LOFAR self-calibration}
Before self-calibrating the target, the phase solutions found on the infield calibrator were applied. Several iterations of self-calibration were subsequently performed in order to obtain the final calibrated data set. These consisted of  phase-only self-calibration, followed by amplitude-and-phase self-calibration. Both were repeated until no significant improvement in terms of signal-to-noise ratio was obtained. This occurred after four phase-only and seven amplitude-and-phase iterations. After each calibration cycle, a new image with a corresponding model was created with WSClean \citep{Offringa2014} using a robust weighting with a robust parameter of $-1$, multi-scale clean, and multi-frequency synthesis (MFS) deconvolution.

For an objective measure of whether or not the image quality had increased, after a cycle of self-calibration we used the absolute ratio of the maximum and minimum pixel value in the image. The maximum value is the peak flux of the target source. The minimum value is most likely due to artefacts around the source from residual errors. An increase in the ratio between the two indicates that there are likely fewer artefacts.

\begin{figure}
        \centering
        \includegraphics[height=0.5\linewidth]{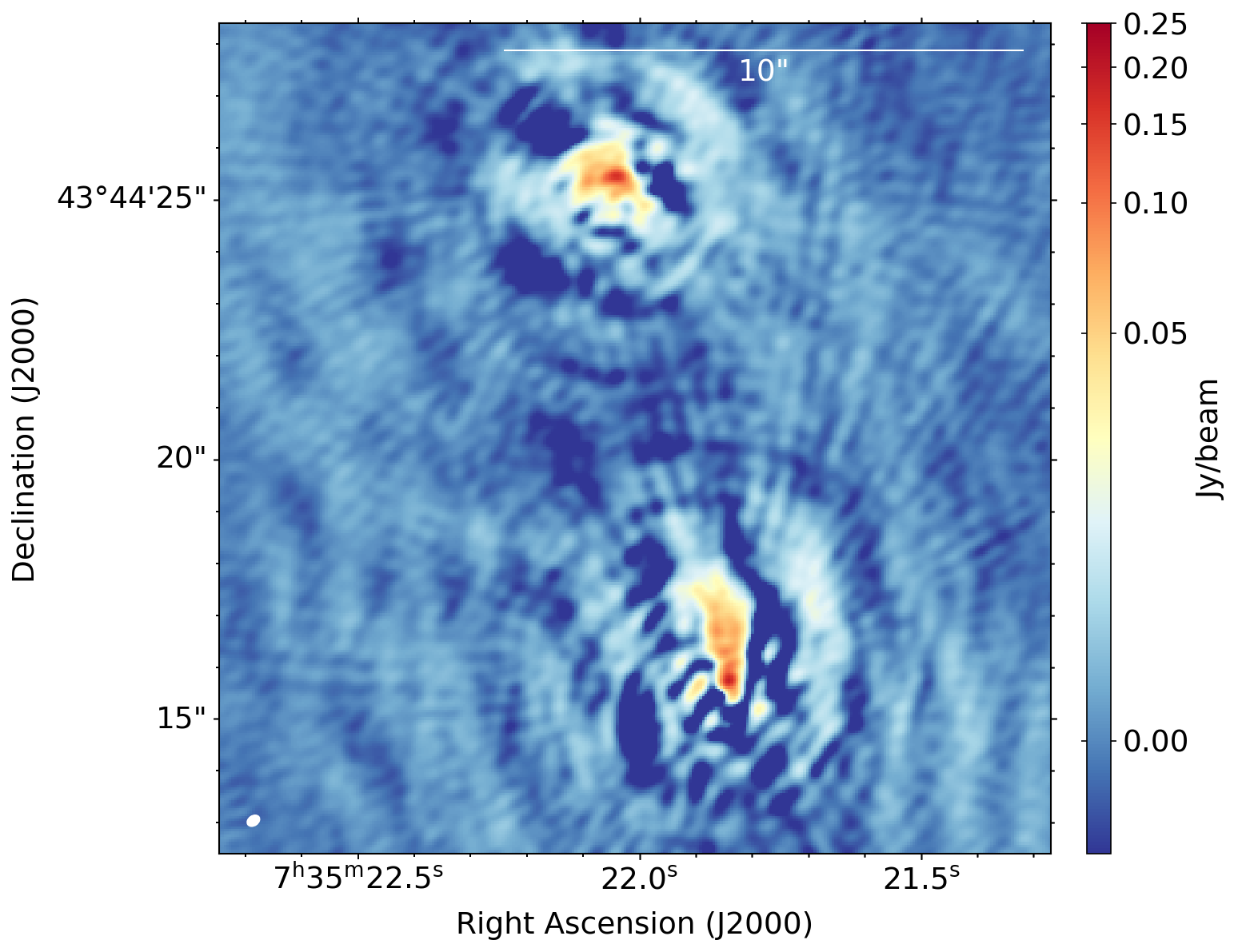}
    \includegraphics[height=0.5\linewidth]{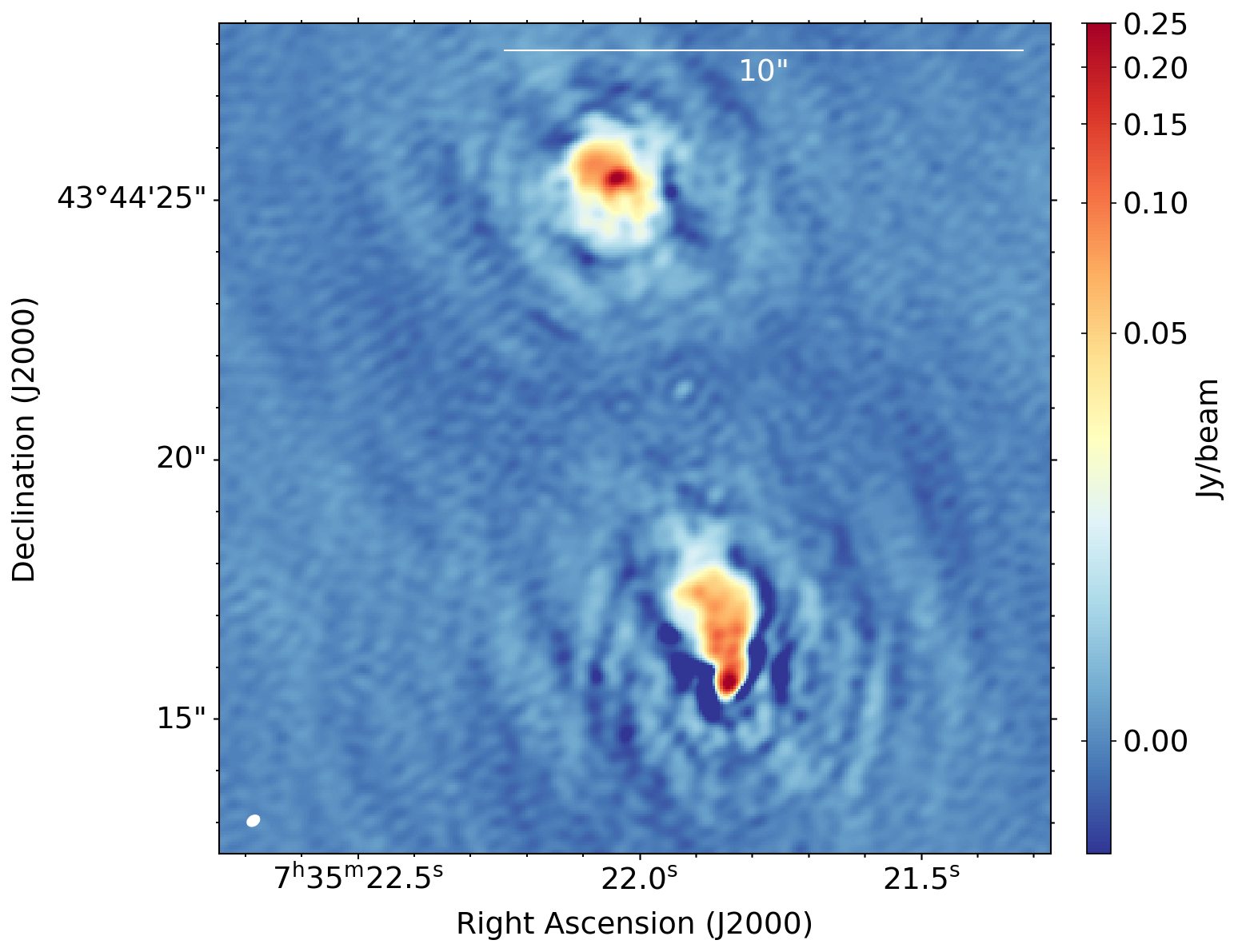}
    \includegraphics[height=0.5\linewidth]{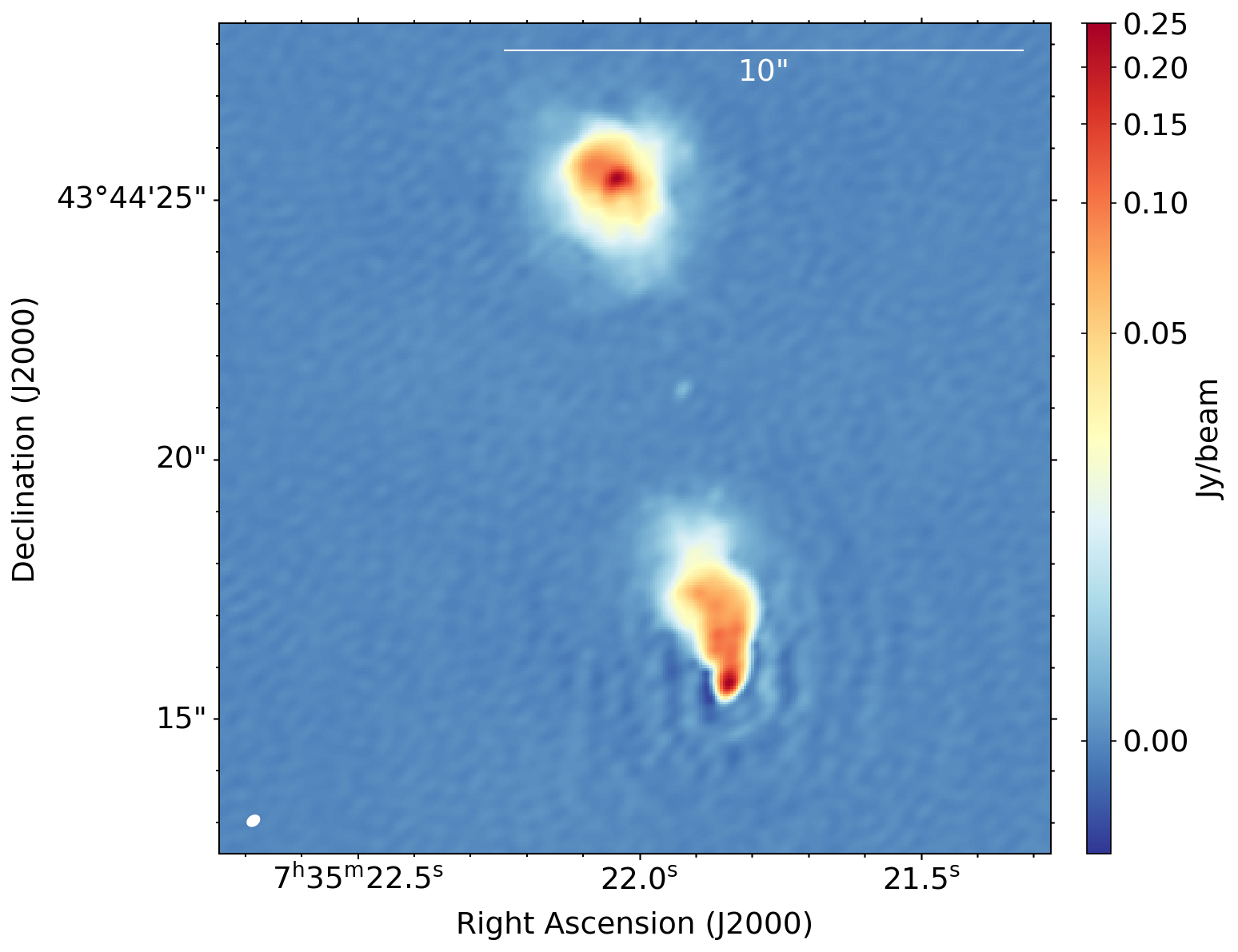}
    \caption{Comparison between the pipeline-calibrated (\textit{top}), phase-only self-calibrated (\textit{middle}), and amplitude-and-phase self-calibrated (\textit{bottom}) images of the $143\ \mathrm{MHz}$ data. The colour scale is shown in an inverse hyperbolic sine ($\mathrm{arcsinh}$) stretch, with the same scale for each image. For each image the pixel scale is $0.05''$. In the bottom left corner the restoring beam is shown (white ellipse), $0.28'' \times 0.22''$ in size. The RMS noise is $172\ \mu\mathrm{Jy}\ \mathrm{beam}^{-1}$ in the final image. North is up and east is to the left.}
    \label{fig:HBA_image}
\end{figure}

\begin{figure*}
        \centering
        \includegraphics[width=0.45\textwidth]{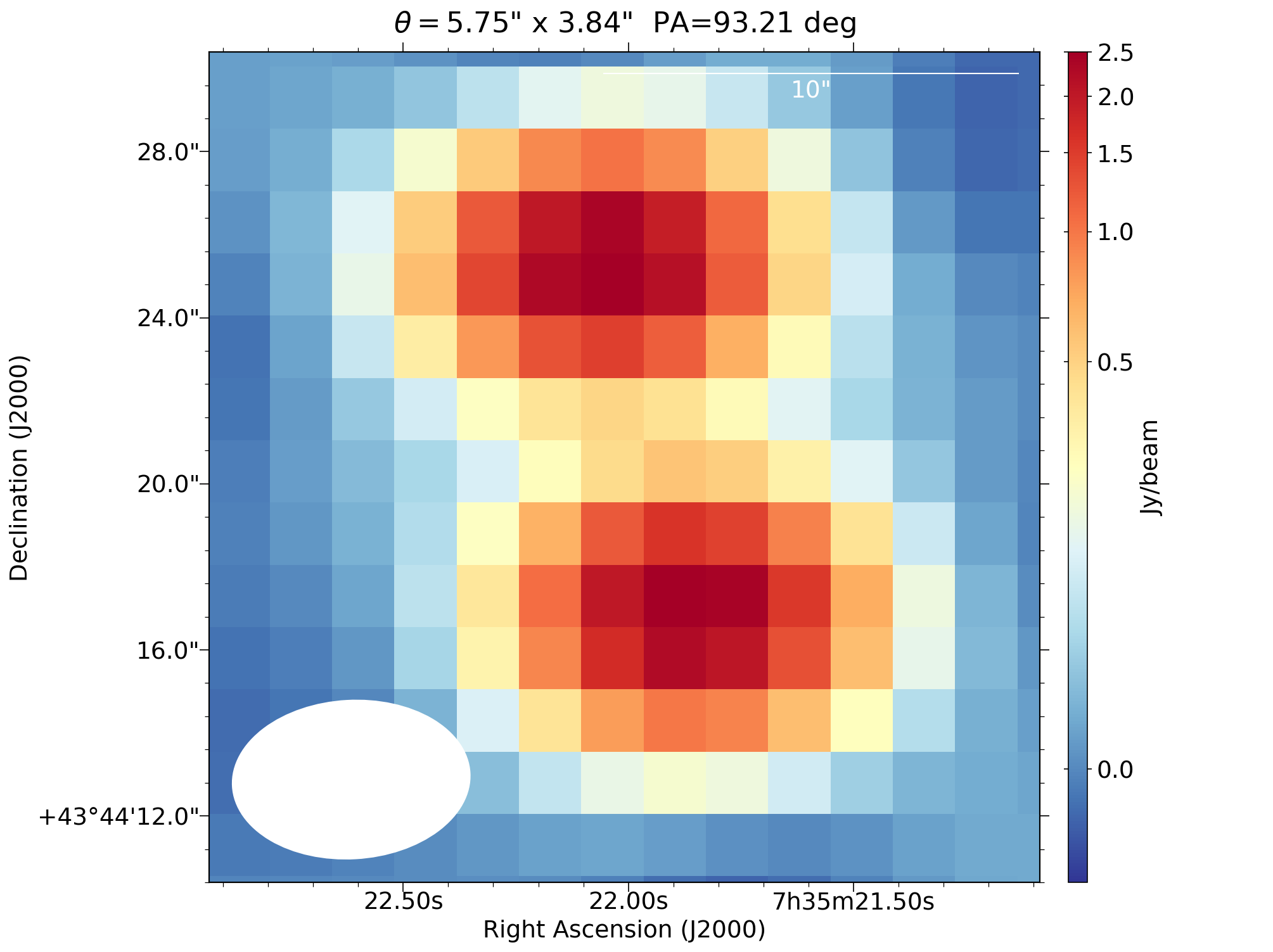}
        \includegraphics[width=0.45\textwidth]{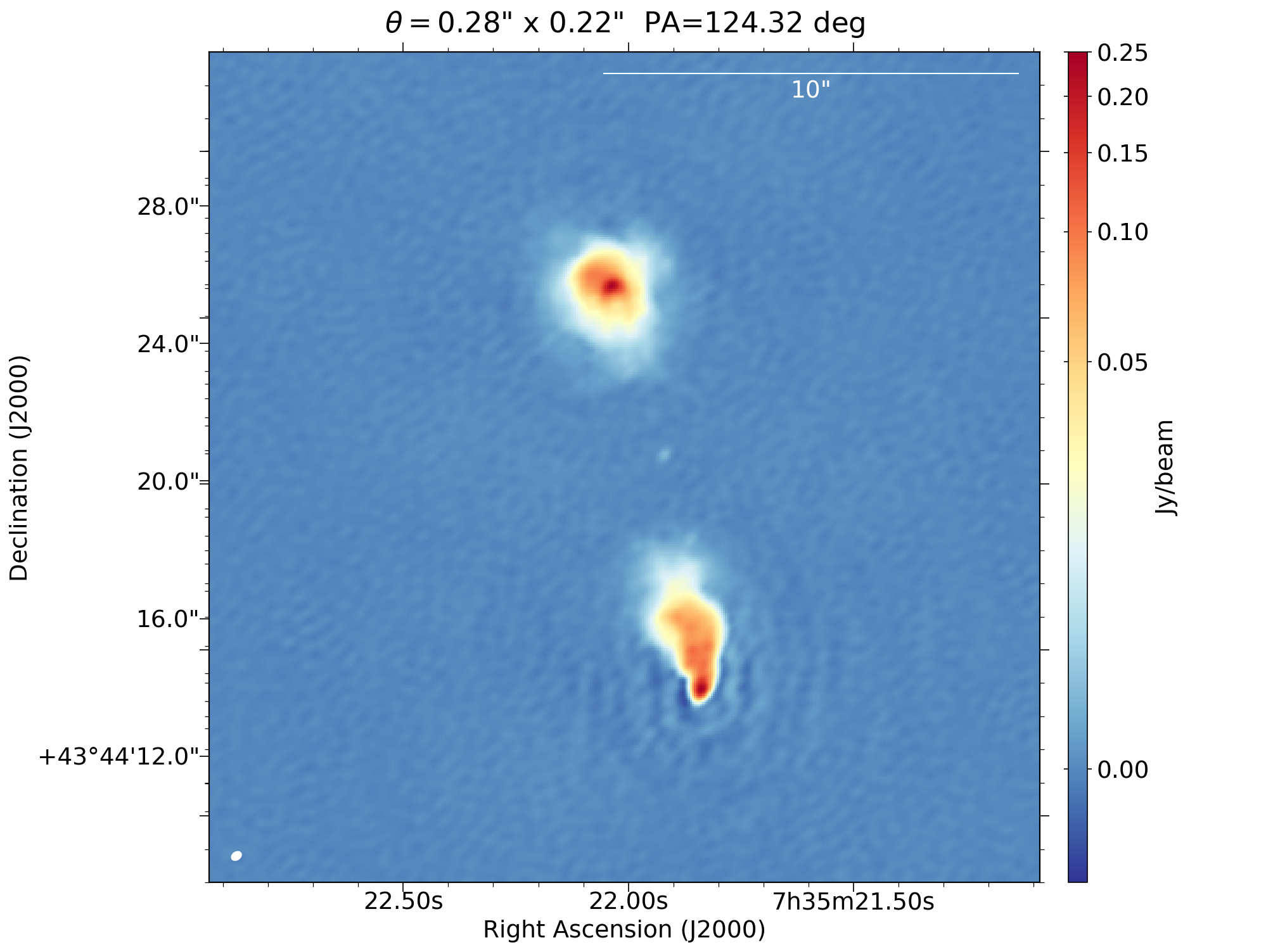}
        \caption{Difference in resolving power at $143\ \mathrm{MHz}$ between using only the Dutch array vs using the full international array. The white ellipse in the bottom left corner shows the beam size, which in each image is $5.8" \times 3.8"$ with a position angle of $93.2\degree$ (\textit{left}) and $0.28" \times 0.22"$ with a position angle of $124.3\degree$ (\textit{right}). In the top right corner a scale bar shows a $10"$ ($83\ \mathrm{kpc}$) reference. North is up and east is to the left. The colour bar indicates intensity, and is shown with a square root stretch.}
        \label{fig:low_high_comparison}
\end{figure*}

The solution interval was gradually reduced during phase-only calibration, starting with a solution for every time slot for the full bandwidth, down to a solution every two time slots and four sub-bands. Gradually tracking faster changes in phase was made possible due to increasing model quality after each iteration. During amplitude-and-phase calibration the solution intervals were kept constant at $\sim 10\ \mathrm{min.}$ and ten sub-bands, which was found to be short enough to allow for time-varying corrections, but not so short as to cause overfitting. The amplitude calibration had an inner uv-cut applied, which was set at $100\ \mathrm{km}$. This improved stability and produced the best results.
We show the progress and final image in Fig.~\ref{fig:HBA_image}. An angular resolution of $0.28'' \times 0.22''$ is obtained. Figure~\ref{fig:low_high_comparison} shows a comparison between an image with only the Dutch array, with a resolution of the order of $\sim 6"$ and an image with the international stations.

\begin{figure*}
        \centering
        \includegraphics[width=0.45\textwidth]{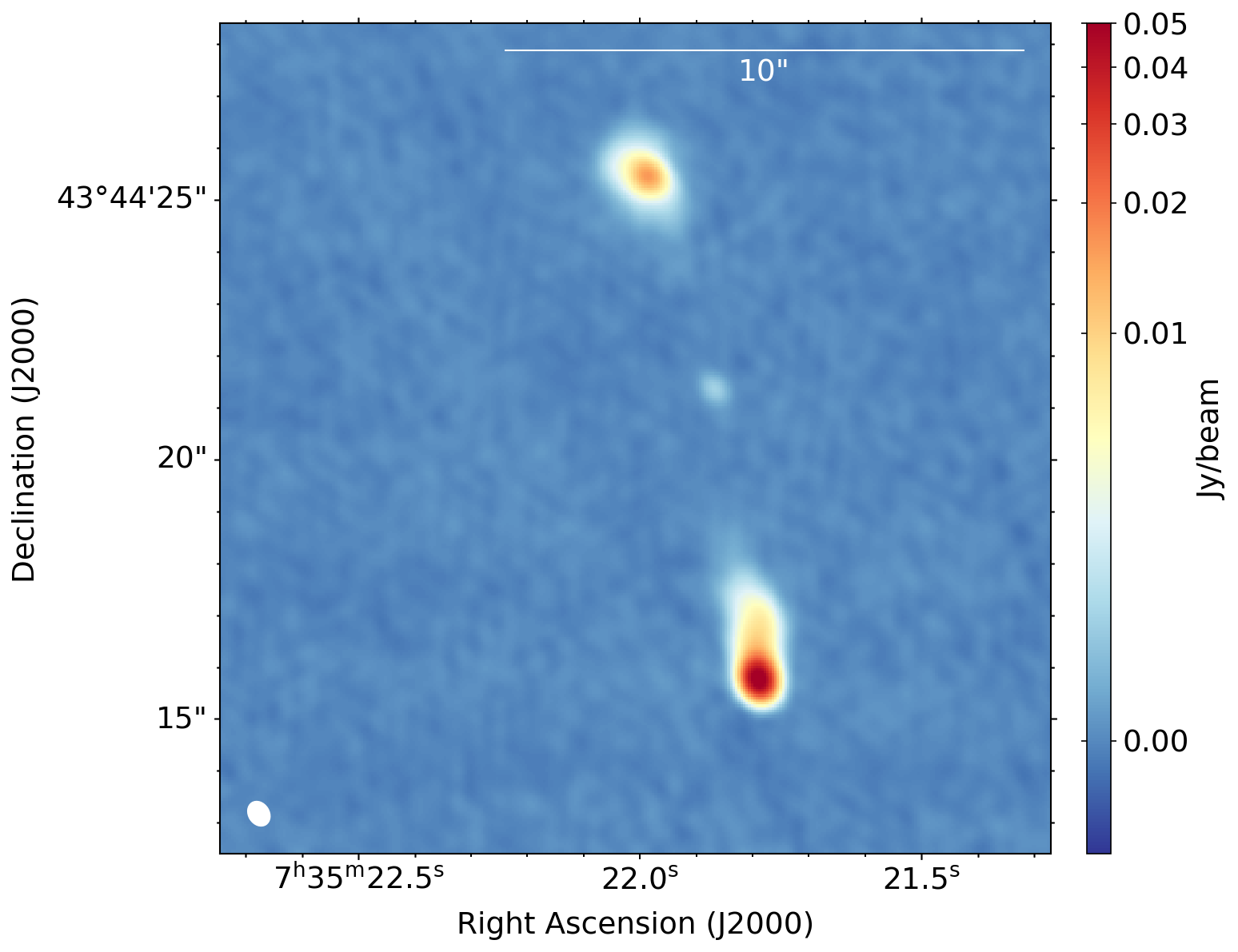}
        \includegraphics[width=0.45\textwidth]{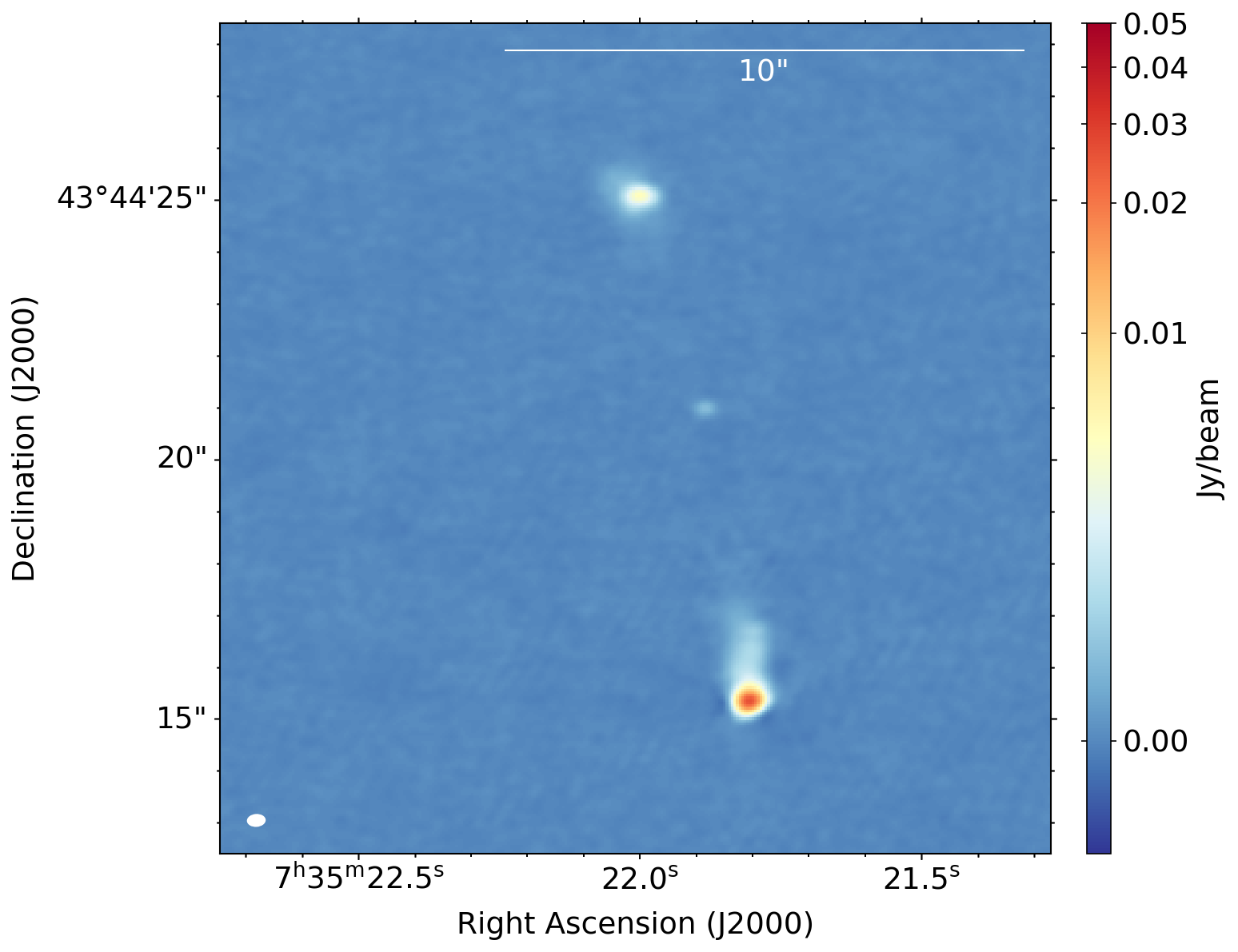}
        \caption{Natural weighted radio maps showing the re-reduced VLA $4.7$ GHz (\textit{left}) and $8.44$ GHz (\textit{right}) data. The colour scale is shown on an inverse hyperbolic sine ($\mathrm{arcsinh}$) stretch from $-0.001$ to $0.05$ Jy beam$^{-1}$. In the bottom left corner the restoring beam is shown as a white ellipse. This is $0.52" \times 0.42"$ with a position angle of $32.2\degree$ and $0.36" \times 0.25"$ with a position angle of $-86.1\degree$, for the $4.7$ and $8.44$ GHz maps, respectively.}
        \label{fig:vla}
\end{figure*}

\subsection{Archival VLA recalibration}
To obtain radio maps at higher frequencies, the archival VLA data, were recalibrated using CASA 5.3.0 \citep{McMullin2007}. The data used here  from the old VLA, so before reprocessing the data were imported into the CASA image format using \texttt{importvla}.

First, the calibrator was determined and its model applied, with \texttt{setjy}, using the \cite{PerleyButler2017} flux scale. For the X-band this was 3C\,48 and for the C-band this was 3C\,286. We then solved for phase-only gains on a one-minute solution interval, followed by an amplitude-and-phase calibration over the entire observation. This was done for both the flux density calibrator and the phase reference using the \texttt{gaincal} task.

Next, the phase reference was put on the correct flux density scale using the \texttt{flux scale} task, transferring the information from the flux density calibrator. We then solved for the gains of the phase reference in the same way as described above. Finally, the corrections were applied to the data, and the data were  imaged using \texttt{clean}. Subsequent  amplitude-and-phase self-calibration concludes the reduction of the archival VLA data. Figure~\ref{fig:vla} shows the resulting radio maps.

\subsection{Resolution matching and astrometric corrections}
For spectral comparison, the maps of 4C 43.15 at different frequencies need to be matched in terms of resolution, uv-coverage, and astrometry. The highest resolution that can be obtained is set by the frequency with the lowest native resolution, in this case the LBA data. Excluding the LBA data, the lowest resolution is set by the HBA data. Table~\ref{tab:reimaging} summarises the reimaging parameters. To match the uv-coverage of LOFAR and VLA an inner cut of $20\ \mathrm{k}\lambda$ was used. This limit is set by the X-band data. Outer cuts of $230\ \mathrm{k}\lambda$ and $589\ \mathrm{k}\lambda$ were used to get close to the target resolutions of $0.9''$ and $0.38"$, respectively, after which the images were further smoothed to the desired resolution using a Gaussian kernel. The CASA \texttt{imsmooth} task was used for smoothing. Finally, the image was rebinned to the desired pixel scale with \texttt{imrebin}. Regridding was needed for the VLA C-band image, from B1950 to J2000. The final products are images with a $0.9''$ circular beam and a $0.2''$ pixel scale, and images with a $0.38''$ circular beam and a $0.05''$ pixel scale.

\begin{table}
        \centering
    \caption{Final reimaging parameters for each data set for the $0.9''$ resolution maps.}
        \begin{tabular}{l|l|l|l}
        Data set & Frequency & Weighting & UV range\\
        \hline\hline
        LOFAR LBA & 55 MHz & robust $-1.5$ & $>1.4\ \mathrm{k}\lambda$ \\
        LOFAR HBA & 143.26 MHz & robust $-1$ & $20 - 230\ \mathrm{k}\lambda$ \\
        VLA (A conf.) & 4.74 GHz & natural & $20 - 230\ \mathrm{k}\lambda$ \\
        VLA (A conf.) & 8.46 GHz & natural & $20 - 230\ \mathrm{k}\lambda$ 
    \end{tabular}
    \label{tab:reimaging}
\end{table}

Except for the LBA map, we aligned the images relative to each other using the compact radio core. In the LOFAR HBA and the VLA data, the core can clearly be seen. Therefore, we corrected the astrometry of these images, using the K-band continuum image of \cite{Motohara2000}. First a Gaussian fit was made to the K-band image to determine the host galaxy's position. This gave us a position of $\alpha_\mathrm{K} = 7\mathrm{h}35\mathrm{m}21.92\mathrm{s}$, $\delta_\mathrm{K} = 43\degree44'20.70''$. The uncertainties on the fit are $1\ \mathrm{mas}$ in both right ascension and declination. The core was then fitted with a Gaussian in all the other images. The centroids of the Gaussians were then subtracted from the centroid of the K band to determine the astrometric offsets, which are summarised in Table~\ref{tab:astrometry}. Using these offsets the images were then aligned in pixel space to the K-band image, using the \texttt{OGEOM} task in AIPS\footnote{\textit{Astronomical Image Processing System}, \url{http://www.aips.nrao.edu/index.shtml}}.

This procedure could not be used for the LBA image because the radio core was not detected in the LBA map. Hence, the alignment between LBA and HBA images was done assuming a Gaussian shape for the lobes.

\begin{table}
        \centering
        \caption{Offsets in right ascension and declination for the astrometric correction of the LOFAR HBA data and VLA C- and X-band data, with respect to the K-band position.}
        \begin{tabular}{r|c|c}
        Frequency & $\Delta$RA [arcsec] & $\Delta$DEC [arcsec] \\
        \hline\hline
        $143.26\ \mathrm{MHz}$ & $0.06$ & $-0.6$ \\
        $4.71\ \mathrm{GHz}$ & $1.8$ & $-0.45$ \\
        $8.44\ \mathrm{GHz}$ & $0.5$ & $-0.3$ \\
        \end{tabular}
        \label{tab:astrometry}
\end{table}

\subsection{Spectral modelling}
Here we describe the procedure that was used for the spectral modelling. Section 4.3 presents the results. We used the \textsc{BRATS} software package to produce spectral index maps and to perform the spectral modelling. The northern and southern lobes were modelled separately because of their apparent difference, but the procedure was the same for both.

The four radio maps were loaded into \textsc{BRATS} together with DS9 region files defining the two lobes and an empty sky region for measuring the RMS noise. The flux density uncertainties for the VLA were set to $11\%$ to be consistent with the measurements of \cite{Carilli1997}, and set to $20\%$ for the LBA \citep{Morabito2016} and HBA frequencies.

The HBA uncertainties are dominated by uncertain beam models that affect the transfer of amplitudes from the flux density calibrator to the target source and uncertainties in the flux density of the calibrator source. An initial cut of $5\sigma_\mathrm{RMS}$ was then made to determine the pixels to use during modelling. The \texttt{signaltonoise} parameter in \textsc{BRATS} allows for  an additional constraint in the signal-to-noise ratio when deciding what pixels to keep. This parameter was set to $3$ for the northern lobe and $6$ for the southern lobe. Values lower than this resulted in \textsc{BRATS} rejecting the model fits. Its values were determined manually and balance the number of pixels used versus the goodness of fit. A detailed explanation of this parameter can be found in \cite{Harwood2013}. Spectral index maps were made using a weighted linear least-squares fit to the data that remained after the above signal-to-noise criteria were applied.

An estimate of the magnetic field is required before any further fitting of spectral models can be done. This estimate is based on the revised equipartition field strength presented in \cite{Beck2005}, who present an alternative formulation of the equipartition formula in terms of a constant proton-electron ratio $K_0$ (which is $0$ for FRIIs; see e.g. the results from \citealt{Hardcastle2002,Croston2004}, and \citealt{Croston2005}), the intensity of the synchrotron radiation and the spectral index. Over the years evidence has been found that the magnetic field can be significantly weaker  than that estimated from equipartition. \cite{Perley1991} found a factor of three difference in 3C 295 when comparing the equipartition field strength with the field strength required to match the hotspot advance speed with the speed at which the lobe and core were moving apart. Similarly, \cite{Wellman1997} found that a ratio of $B/B_\mathrm{eq} = 0.25$ was required for consistent interpretation of their data, and that the general scatter around this value in their sample was small ($\sim15\%)$. \cite{Croston2005}  used X-ray observations of 33 radio galaxies and quasars and found that it
is common for magnetic field strengths to be different from those calculated
via equipartition, and \cite{Harwood2016} confirmed a departure of equipartition for 3C 452 and 3C 223 showing that it may be the cause of discrepancies between the spectral and dynamical ages. Recently, \cite{Ineson2017} studied a sample of $47$ radio galaxies at both X-ray and radio wavelengths, and \cite{Turner2018} studied sources in the Third Cambridge Catalogue of Radio Sources (3C). Both authors find a median magnetic field strength of $B = 0.4B_\mathrm{eq}$. \cite{Mahatma2019} also identify the range $B = 0.1-0.5 B_\mathrm{eq}$ as plausible. We therefore scale our equipartition magnetic field strength by $0.4$.

A search for the best fitting injection index was then performed using the \texttt{findinject} task in \textsc{brats}, and finally a spectral ageing model was fitted to the data. \textsc{brats} allows  different models to be fitted: Kardashev-Pacholczyk (KP; \citealt{Kardashev1962}), Jaffe-Perola (JP; \citealt{Jaffe1973}), or  Tribble (\citealt{Tribble1993}). The difference between the KP and JP model is that in the JP model, the electron pitch angle is an average value over its radiative lifetime instead of a constant value. The Tribble model describes the magnetic field as a Gaussian random field instead of a constant value. 4C 43.15 was fitted with a JP+Tribble model, which is thought to be the most realistic of the three because it allows the magnetic field to vary across the source \citep{Hardcastle2013,Harwood2013}. First, the injection index $\alpha_\mathrm{inj}$ was determined through a coarse search between $-0.5$ and $-1.0$, with steps of $0.05$. This yields a best fit in an interval $\Delta\alpha_\mathrm{inj}=0.05$ wide. Another finer search was then done in this interval, with a step size of $0.01$. Finally, the \texttt{fitjptribble} task was used to fit the ageing model.

\section{Results}
\subsection{A GHz steep-spectrum core}
With the core detected at $143\ \mathrm{MHz}$, $4.71\ \mathrm{GHz,}$ and $8.44\ \mathrm{GHz}$, its overall spectral shape can be determined. Typically, radio galaxy cores have flat spectra in the GHz regime. It has been shown, for example by \cite{Athreya1997} and \cite{Carilli1997}, that the cores of high-redshift radio galaxies can have spectral indices $\alpha < -0.5$, between $4.7\ \mathrm{GHz}$ and $8.3\ \mathrm{GHz}$, including the core of 4C 43.15.

Flux densities were measured, using $0.38''$ matched resolution maps, using the same aperture for all three images, based on the $3\sigma_\mathrm{RMS}$ contours of the HBA map, where $\sigma_\mathrm{RMS}$ is the RMS map noise. The measured flux densities are listed in Table~\ref{tab:core}.

The spectral indices between these frequencies are $\alpha_{4.71\ \mathrm{GHz}}^{8.44\ \mathrm{GHz}} = -0.83 \pm 0.04$, $\alpha_{8.44\ \mathrm{GHz}}^{143\ \mathrm{MHz}} = -0.35 \pm 0.05$, and $\alpha_{4.71\ \mathrm{GHz}}^{143\ \mathrm{MHz}} = -0.27 \pm 0.06$. Uncertainties were derived using Gaussian error propagation, neglecting the uncertainty in the effective frequency at which the measurements were taken.

\begin{table}
        \centering
        \caption{Flux densities of the radio core, measured using the same aperture, in $0.38''$ matched resolution maps. Uncertainties are taken to be $20\%$ for the HBA and $2\%$ for the VLA,  the same as for the spectral modelling.}
        \begin{tabular}{l|l|l}
        Frequency & $S_\mathrm{int}$ [mJy] \\
        \hline\hline
        $143\ \mathrm{MHz}$ & $3.0 \pm 0.6$ \\
        $4.71\ \mathrm{GHz}$ & $1.2 \pm 0.02$ \\
        $8.44\ \mathrm{GHz}$ & $0.72 \pm 0.01$ \\
        \end{tabular}
        \label{tab:core}
\end{table}

\subsection{ILT radio morphology and measurements}
From the intensity map it is clear that the two lobes are different. The northern lobe has a more circular morphology as projected on the plane of the sky. It does not show a sharp termination of the hotspot where it meets the surrounding medium. Instead, its hotspot appears to be embedded in a region of diffuse emission. The southern lobe has a more conical shape with a clear termination between hotspot and its surrounding medium, showing no diffuse emission past this point.
The bridge of emission seen at $55\ \mathrm{MHz}$ is not seen at $143\ \mathrm{MHz}$ or higher frequencies. No jet emission is identified either, although a small patch of emission just slightly north-east of the core can be identified at the $3-5\sigma_\mathrm{RMS}$ level. Table~\ref{tab:lobeflux} presents the flux densities of both lobes, measured from the HBA and reprocessed VLA maps.
The reprocessed VLA maps are on average $30\%$ brighter than the flux densities reported by \cite{Morabito2016}, but are within $11\%$ of the values reported by \cite{Carilli1997} in terms of total flux density.
The flux densities of the northern and southern lobes are equal within the uncertainties at $143\ \mathrm{MHz}$. Imaging the source with only the Dutch array yields an integrated flux density of $S = 6.2 \pm 1.2\ \mathrm{Jy}$ (assuming a $20\%$ uncertainty). This is consistent with scaling the 6C value of $5.9\ \mathrm{Jy}$ \citep{Hales1993} from $151\ \mathrm{MHz}$ to $143.26\ \mathrm{MHz}$ using the measured spectral index of $\alpha = -0.83$ \citep{Morabito2016}. The total integrated flux density presented in Table~\ref{tab:lobeflux} (measured inside $5\sigma_\mathrm{RMS}$ contours) is lower than the measurement from the low-resolution map, but is consistent within the uncertainties. When using the same aperture as for the low-resolution map, however, an integrated flux density of $S = 6.3 \pm 1.3\ \mathrm{Jy}$ (assuming a $20\%$ uncertainty) was found. This likely indicates that  some of the flux density is still contained in the artefacts such as those around the southern lobe.

The $143\ \mathrm{MHz}$ flux density of 4C 43.15 corresponds to a luminosity of $L_\mathrm{143 MHz} = 2.3 \times 10^{29}\ \mathrm{W}\ \mathrm{Hz}^{-1}$;  the luminosity is given by

\begin{equation}
        L_\nu = 4\pi D_L^2 S_\nu (1 + z)^{-(1 + \alpha)},
\end{equation}
where $D_L$ is the luminosity distance, $S_\nu$ is the flux density, $z$ is the redshift, and $\alpha$ is the spectral index. The integrated spectral index of $\alpha = -0.83$, for frequencies below $500\ \mathrm{MHz}$ (see \citealt{Morabito2016}), was used to calculate $L_\nu$.

In the $143\ \mathrm{MHz}$ map the angular size of the source from the northern hotspot to the southern hotspot is $9.9''$ on the sky, corresponding to a linear size of $82\ \mathrm{kpc}$. The positions of each hotspot and the core were determined by fitting Gaussian functions using the \texttt{CASA} viewer. These positions were then used to determine the parameters listed in Table~\ref{tab:lobeflux}. Here the jet angle is defined as the angle anti-clockwise with respect to the nearest vertical at the position of the radio core.

\begin{table*}
        \centering
        \caption{Flux densities of the northern and southern lobes, measured from native resolution maps. All are measured within $3\sigma_\mathrm{RMS}$ contours, except for LOFAR, which was measured within $5\sigma_\mathrm{RMS}$. Quoted errors in flux density are the combination of calibration uncertainty and the RMS noise in the map: $\sqrt{\sigma_\mathrm{map}^2 + \sigma_\mathrm{cal}^2}$. The errors in core distance and jet angle come from uncertainties from fitting and uncertainty propagation. Reported jet angles are taken as east of north, modulo $90\degree$ (i.e. with respect to their closest vertical).}
        \begin{tabular}{c|l|l|l|l|l}
        & Frequency & $S_\mathrm{int}$ [mJy] & RMS [mJy beam$^{-1}$] & Core distance [kpc] & Jet angle [$\degree$]\\
        \hline\hline
          & $143.26\ \mathrm{MHz}$  & $(3.0 \pm 0.6)\times 10^3$ & $0.17$ & $35.4 \pm 0.2$ & $17.4 \pm 6.5\ 10^{-2}$\\
        N & $4.71\ \mathrm{GHz}$ & $39.6 \pm 4.4$ & $0.072$ & $35.9 \pm 0.2$ & $17.8 \pm 6.0 \ 10^{-2}$\\
          & $8.44\ \mathrm{GHz}$ & $15.4 \pm 1.7$ & $0.034$ & $35.6 \pm 0.1$ & $17.3 \pm 4.5\ 10^{-2}$\\
        \hline
          & $143.26\ \mathrm{MHz}$  & $(3.0 \pm 0.6) \times 10^3$ & $0.17$ & $47.2 \pm 0.1$ & $9.0 \pm 1.5\ 10^{-2}$\\
        S & $4.71\ \mathrm{GHz}$ & $113.2 \pm 12.5$ & $0.072$ & $46.6 \pm 0.1$ & $8.6 \pm 1.3\ 10^{-2}$ \\
          & $8.44\ \mathrm{GHz}$ & $56.1 \pm 6.2$ & $0.034$ & $47.2 \pm 0.1$ & $8.6 \pm 1.5\ 10^{-2}$ \\
          \hline
        \end{tabular}
        \label{tab:lobeflux}
\end{table*}

\subsection{Spectral index maps and model fitting}
\begin{figure*}
        \includegraphics[width=\textwidth]{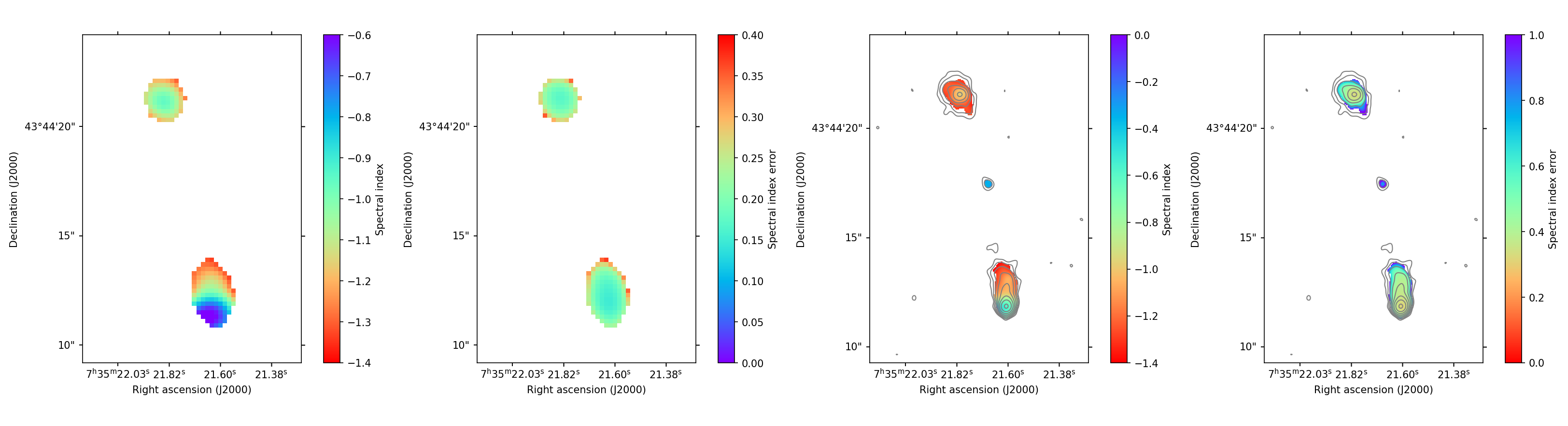}
        \caption{Spectral index maps of 4C43.15 obtained from BRATS and considering only pixels with values $>5\sigma_\mathrm{RMS}$. \textit{Left:}  $0.9''$ resolution map made from the LBA, HBA, and VLA data. The same additional signal-to-noise constraint as when fitting the ageing model was used. \textit{Right:}  $0.38''$ pixel-by-pixel map made from the HBA and VLA data. No additional signal-to-noise constraint was used in BRATS. The grey contours are drawn based on the $4.71\ \mathrm{GHz}$ map, starting at $0.29\ \mathrm{mJy\ beam}^{-1}$ and $0.38\ \mathrm{mJy\ beam}^{-1}$ for the left and right images, respectively. Each contour represents a factor of two increase in intensity.}
        \label{fig:specindex}
\end{figure*}

Figure~\ref{fig:specindex} shows the spectral index maps obtained. The map including the LBA data has a resolution of $0.9''$; the map excluding the LBA map has a resolution of $0.38''$. As expected, the spectra become steeper in directions away from the hotspots and towards the lobes. In the northern lobe the spectral index ranges from $-1.0 \pm 0.2$ to $-1.3 \pm 0.4$ and in the southern lobe from $-0.6 \pm 0.2$ to $-1.3 \pm 0.3$. For the high-resolution map at $0.38''$ the spectral index ranges from $-1.0$ to $-1.4$ in the northern lobe and up to $-1.5$ in the southern lobe.

The high-resolution map shows a steep drop in the spectral index in the   south-western corner of the southern lobe, which is likely due to the remaining negative artefact around the source. This prevents a reliable lower value of the spectral index from being determined. The core has an average spectral index of $-0.3$, consistent with earlier calculations of its integrated spectral index.

The spectral index map clearly shows a difference between the northern and southern lobes of 4C43.15. The southern lobe shows a  wide range of spectral indices ($\Delta\alpha = 0.85$) between the flattest and steepest parts. The northern lobe shows a narrower spread in spectral indices ($\Delta\alpha = 0.39$), and is steeper on average.

\begin{figure*}
        \includegraphics[width=\textwidth]{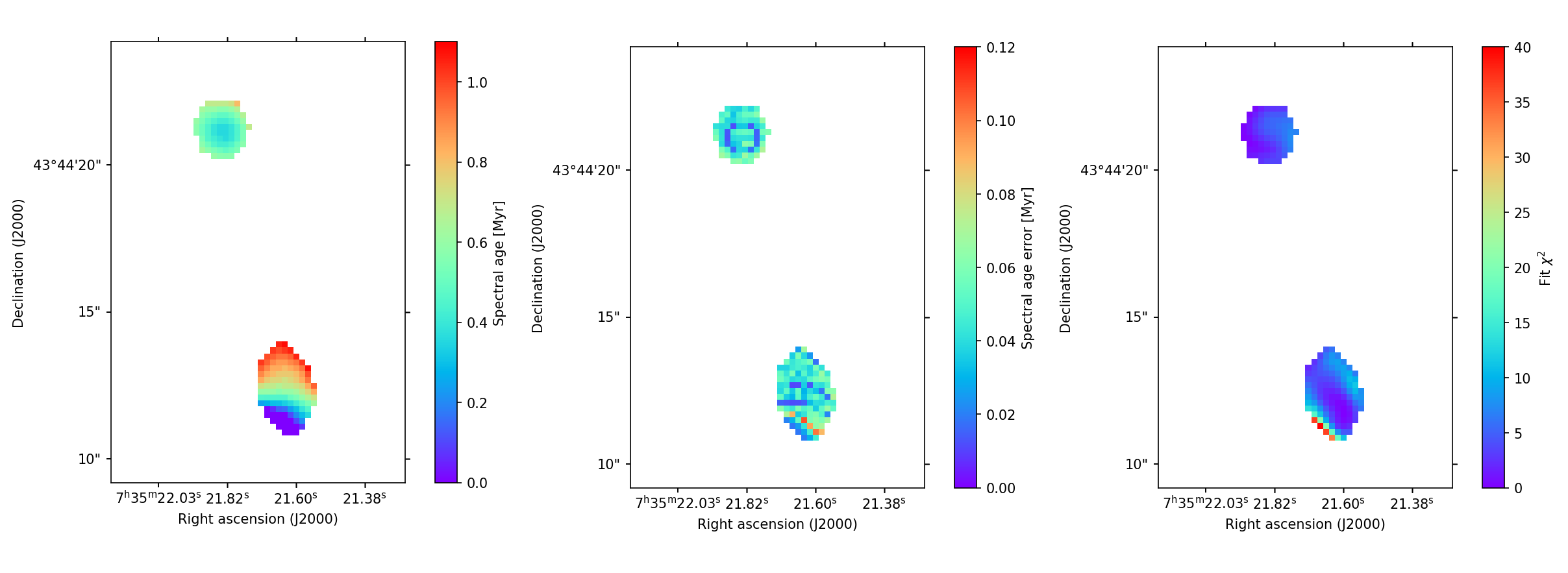}
        \caption{Spectral age maps of the fitted regions in 4C 43.15 after fitting a JP+Tribble model to the LBA, HBA, and VLA data. From left to right the panels show the spectral age of the plasma, its positive uncertainty, and the $\chi^2$ value of the regions. The maps are shown at a $0.9''$ resolution. The grey contours are drawn based on the $4.71\ \mathrm{GHz}$ map, starting at $0.29\ \mathrm{mJy/beam}$. Each contour represents a factor of $\sqrt 2$ increase in intensity.}
        \label{fig:specage}
\end{figure*}

Assuming equipartition, a magnetic field strength of $B_\mathrm{eq} = 13.1\ \mathrm{nT}$ was estimated. Scaled by the previously discussed factor  of $0.4$, this translates to an actual magnetic field strength of $B = 5.2\ \mathrm{nT}$ for the lobes. Because of their different injection indices, the northern and southern lobe were modelled separately. A best fit for the injection index of $\alpha^\mathrm{north}_\mathrm{inj} = -0.8$ and $\alpha^\mathrm{south}_\mathrm{inj} = -0.6$ was found for the northern and southern lobes, respectively. Figure~\ref{fig:injectindex} shows the $\chi^2$ landscape of the explored injection indices. In Fig.~\ref{fig:specage} we present the spectral age maps, obtained by fitting a JP+Tribble model to the $0.9''$ data. For the northern lobe the model could not be rejected at the $90\%$ significance level, and spectral ages in the range $\tau^\mathrm{min}_\mathrm{spec} = 0.36^{+0.04}_{-0.06}\ \mathrm{Myr}$ to $\tau^\mathrm{max}_\mathrm{spec} = 0.81^{+0.05}_{-0.06}\ \mathrm{Myr}$ were found. For the southern lobe the model could not be rejected at the $90\%$ significance level, and spectral ages in the range $\tau^\mathrm{min}_\mathrm{spec} = 0.0^{+0.02}_{-0.0}\ \mathrm{Myr}$ to $\tau^\mathrm{max}_\mathrm{spec} = 1.08^{+0.07}_{-0.05}\ \mathrm{Myr}$ were found. These ages result in hotspot advance speeds of $0.15c$ and $0.14c$, respectively, for the northern and southern hotspot. Here $c$ is the speed of light. Figure~\ref{fig:modelfits} shows model fits to the youngest, oldest, and best fitting regions of each lobe.

\section{Discussion}
With the high angular resolution of the ILT, spectral studies can now be done on distant objects. For example, \cite{Harris2019} studied jet knots in 4C\,41.19 and they favour an intrinsically curved particle energy distribution as the cause behind the observed spectral curvature in the knots. The spectral modelling presented in this work focuses on lobes, and is similar to the recent modelling in \cite{Harwood2016, Harwood2017}. 

The northern lobe has a rounder shape and  the hotspot is embedded in the lobe, while the southern lobe is more elongated with the hotspot sitting at the edge. This could be a consequence of the ambient density being almost a factor of two lower for the northern lobe, as determined by \citet{Motohara2000} from observations of the $H\alpha$ emission line. This would allow the lobe to expand more easily into its surroundings. The southern lobe is facing a higher ambient density. Material here will thus have a harder time expanding into its surroundings, taking on a more cylindrical shape. It could also be more simply an orientation effect. There is no estimate of the inclination of the source available, which can complicate morphological analysis, but overall the morphology of 4C 43.15 is that of a typical FRII radio galaxy.

The closest example of such an object is the well-studied radio galaxy Cygnus A (e.g. \citealt{Carilli1991,Steenbrugge2010,McKean2016,DeVries2018}; hereafter Cyg A). We therefore first compare our findings for 4C 43.15 with it and two other nearby sources studied in \cite{Harwood2016} and \cite{Harwood2017}: 3C 223 and 3C 452, $z=0.137$ and $z=0.081$, respectively. We see the expected morphological similarities between the two sources, but differences are also seen. In Cyg A two hotspots are identified in each lobe, which may be attributed to the jet changing direction over time \citep{Carilli1999}. In 4C 43.15 there is a smaller extent of diffuse emission from material flowing back to the core compared to the other sources. It reaches about midway between the hotspot and the core, whereas the nearby sources have diffuse emission close to the core. This could indicate that the material has not had the time to flow back yet, reflected by the two orders of magnitude difference in spectral age.

The hotspots in 4C 43.15 are advancing faster than those in these local sources. Estimates for Cyg A range from $0.005c$ to $0.05c$ \citep{Muxlow1988,Alexander1996} and for 3C 223 and 3C 452 they are between $0.01c-0.02c$, which is  an order of magnitude slower. 4C 43.15 has a significantly higher radio luminosity than these local sources, and hence a more powerful jet. The kinetic power of the jets increases with radio luminosity \citep{Hardcastle2019}. A more powerful jet could thus produce faster moving hotspots, being less affected by the environmental density, due to a larger driving force pushing material out.

These are only three local sources however. To put 4C 43.15 in a broader context, we compare its advance speeds and injection indices with other radio galaxies in the literature. \cite{AlexanderLeahy1987}, \cite{Liu1992}, and \cite{Odea2009} measured spectral indices for a number of sources over a range of redshifts, and estimated advance speeds and spectral ages based on their estimated break frequencies. These ages depend on the adopted ratio $B / B_\mathrm{eq}$, but are between ${\sim} 10^6$ and a few times $10^7\ \mathrm{Myr}$, comparable to what we find for this source. The corresponding advance speeds range up to $0.4c$. They find injection indices between $-0.65$ and $-1.11$. We consider our values to be consistent with those found in the literature.

Finally, the spectral index and spectral age maps show the same characteristics for all sources: relatively young and flat emission at the hot spots, steepening and growing older towards the core. For 3C 223 and 3C 452 the spectral index ranges from $\alpha \approx -0.5$ in the hotspots to $\alpha \approx -1$ in the oldest regions near the core. What makes 4C 43.15 different is a steeper spectrum in general, and the difference between the two lobes. This asymmetry between the lobes is discussed next.

\subsection{What drives the lobe difference?}
The intensity map and spectral index map both show the lobes to be different. The northern lobe has a more diffuse circular shape and has steeper spectra. Here we revisit this difference.

We rule out that the difference in steepness is caused by Doppler boosting. \cite{Humphrey2007} calculated the expected spectral index asymmetries from Doppler boosting for various hotspot advance speeds. An advance speed of $0.2\ c$, would only imply a $0.12$ asymmetry in the spectral index of the hotspots for an inclination angle of $50\degree$, much less than the observed difference.

This brings us back to the question of whether the environment it resides in has an effect on steepening the spectrum, specifically the scenario where a higher ambient density would cause a steeper spectrum. The two arguments for this scenario  introduced earlier are  longer confinement of the lobes increasing surface brightness of the oldest emission, and  slower speeds of the plasma near the hotspot leading to a steeper injection index. Estimates of the environmental density are available from \citet{Motohara2000}, who measure the medium surrounding the northern lobe to only have $55\%$ the density of that surrounding the southern lobe. If the steep spectrum of 4C 43.15  indeed comes from these environmental effects, we expect  the steepest spectral index in the southern lobe and  the steepest hotspot spectral index in the southern lobe.

For the first case we measure no difference between the steepest spectral indices in each lobe. The lobe in the less dense environment is also the steepest on average, opposite to what was expected if confinement by a dense medium drove the steepening. The second case provides a stronger argument. A jet working against a denser medium is proposed to have a steeper injection index. In this work we  estimated the spectral index near the hotspot directly from the data and fitted for the injection index. Both measurements show the northern hotspot to have the steepest spectral index, even though it resides in the lower density environment. For the southern lobe the hotspot spectral index and fitted injection index are consistent with the range of measurements in the literature. They are also close to the theoretical lower limit of $\alpha_\mathrm{inj} = -0.5$ \citep{Athreya1998}. We rule out environment as the cause for a steeper spectral index based on the new information about the spatial distribution of spectra in this source.

Having ruled out Doppler boosting based on the estimated advance speeds, and ruled out the effects on the particle acceleration due to the environmental density based on the spatial distribution of spectra, we remain at the original conclusion by \cite{Morabito2016}. It is likely that the difference between the lobes is caused by adiabatic expansion having a stronger effect on the northern lobe due to its lower ambient density. The expansion will lower the surface brightness of the lobe, as is observed. Furthermore, adiabatic expansion will shift the emission by equal amounts in intensity and frequency \citep{Katz-Stone1997}. As the intensity lowers due to expansion, the emission is shifted to lower frequencies. The spectrum will therefore be artificially steepened by this effect, but it is difficult to quantify by how much it has been steepened.

\subsection{Core spectral index}
The spectral index of the core between $4.71\ \mathrm{GHz}$ and $8.42\ \mathrm{GHz}$ that we find is a factor of two less steep than originally found in \cite{Carilli1997}, for  this source and for their sample. It was mentioned, however, that in this case the identification of the core  was uncertain due to  source confusion, a low signal-to-noise ratio, or a core with $\alpha \leq -1$. Our results seem to exclude the last. Our measured spectral index of $-0.8$ is broadly in line with the values reported by \cite{Athreya1997}.

\subsection{Magnetic field strength}
The uncorrected equipartition magnetic field strength of $13.1\ \mathrm{nT}$ found for 4C 43.15 appears to be higher than its local counterparts, even when scaled  by $0.4$. A typical field strength for local counterparts is   $\sim 1\ \mathrm{nT}$. \cite{Krolik1991} show that the equipartition magnetic field in the lobes scales as $B \propto (1 + z)^{0.8}$, assuming synchrotron radiation as the dominant energy loss process and that the distributions of lobe volume and the time electrons spend  here do not scale with redshift. At $z = 2.429$ this scaling could increase the magnetic field strength by a factor of $2.68$. Locally Cygnus A  has an equipartition field strength of $4\ \mathrm{nT}$ \citep{DeVries2018} and the nearby radio galaxy 3C 277.3 (Coma A, $z = 0.0857$, \citealt{Bridle1981}) has a magnetic field strength of $3 - 5\ \mathrm{nT}$ in its southern lobe when assuming equipartition. Scaling both of these nearby galaxies according to the above relation yields similar field strengths to what is observed here. \cite{Chambers1990} also derived similarly high magnetic field strengths for components in the $z=3.8$ radio galaxy 4C 41.17. The field strength values reported by \cite{Liu1992} and \cite{Odea2009} are    similar. Therefore, the strong magnetic field in 4C 43.15 appears to be in line with expectations.

What does its magnetic field strength imply? The  equivalent magnetic field strength of the CMB is $3.7\ \mathrm{nT}$ at this redshift ($B_\mathrm{CMB} = 0.318 (1 + z)^2$; \citealt{Harwood2013}). This is close to  the scaled magnetic field strength value of $5.2\ \mathrm{nT}$ found for 4C 43.15. In this case synchrotron cooling appears to still be the dominant effect of energy losses in the lobe plasma, but only by a factor of $1.4$. At redshifts $z \gtrsim 3$, the CMB's equivalent magnetic field strength becomes $5.2\ \mathrm{nT}$. Given its luminosity, 4C 43.15 is one of  the rare powerful radio-loud AGNs over a range of redshifts up to $z=5$ \citep{Hardcastle2019}. For the assumed scaling of $B = 0.4 B_\mathrm{eq}$, we would therefore expect a significant fraction of sources above redshift $3$ to be dominated by IC losses. Closer sources will still be dominated by synchrotron losses, depending on their magnetic field strength. The proportionality constant between then equipartition magnetic field strength and the actual magnetic field strength spans a wide range of values. A larger sample of sources with an estimate of their magnetic field strength is needed over a wide range of redshifts for any strong conclusions to be drawn.

\subsection{The $\alpha-z$ correlation}
The first question we address here, in the context of the $\alpha-z$ correlation, is whether or not 4C 43.15 is a true ultra-steep spectrum source, which are typically said to have $\alpha < -1$. The key question  is whether the 4C 43.15 spectrum is intrinsically steep, or if it is just observed to be. \cite{Morabito2016} already pointed out that its integrated spectral index below $1.4\ \mathrm{GHz}$ (i.e. $4.8\ \mathrm{GHz}$ rest frame) would no longer classify as  ultra-steep because of a break in the spectrum. This is reflected by the values of the injection index for both lobes as well. The injection index probes the low-frequency power-law part of the spectrum, and does not show ultra-steep values for either lobe. Hence, although it fits the ultra-steep spectrum criteria of \cite{DeBreuck2000} ($\alpha^{325\mathrm{MHz}}_{1.4\mathrm{GHz}} < -1.3$), the spectral index is not ultra-steep across the entire spectrum. Most importantly, the estimated injection indices are in agreement with typical values found for this kind of sources in the literature. This rules out that the spectrum is intrinsically steep, and means instead that it has steepened because of spectral ageing effects.

What does 4C 43.15 therefore tell us about the $\alpha-z$ correlation? The goal was to see if this source could tell us something about the main driving force behind this correlation. Earlier, three possible causes for this relation were introduced: environmental effects, observational bias, and IC losses. \cite{Klamer2006} discuss how a higher density environment could lead to steeper spectra. Based on the scenarios there, if the steep spectral indices found in this source were  driven by environmental density effects, they would be expected in or near the southern hotspot where the ambient density is highest. However, in this case we find that the steeper spectrum belongs to the lobe residing in the lower density environment. Environmental effects are certainly at play. The northern lobe will be more strongly affected by adiabatic expansion because of the significantly lower density around it compared to the southern lobe.

Secondly, could we be affected by an observational bias? As explored by \cite{Ker2012} and other authors, observational biases should be considered carefully and make interpreting correlations such as $\alpha-z$ difficult. From the luminosity of 4C 43.15 it is clear that we are looking at one of the rarer bright sources in the radio sky. The radio luminosity and jet power of a source are related to each other. Work by  \cite{Ineson2017} and \cite{Hardcastle2019}, among others, show increasing jet powers for increasing $150\ \mathrm{MHz}$ luminosities, and these relations put 4C 43.15 at high jet powers. \cite{Blundell1999} presented arguments for a relation between jet power and spectral index, where more powerful jets can create a stronger magnetic field in the hotspots. As a consequence, more powerful radio sources with stronger magnetic fields age more rapidly, producing steeper spectra. If a stronger jet also means a stronger lobe magnetic field strength, we could be affected by the Malmquist bias here, possibly looking at one of the more luminous members of the high-redshift population. A single source is not enough to draw any conclusions in this regard, but this is not an unreasonable assumption: 3C and 4C sources are powerful radio galaxies. Results for 3C sources from \cite{Vaddi2019} support this; in their sample the highest redshift sources also show the highest magnetic field strength.

Finally, we suggest that there is an interaction with CMB photons. Inverse Compton losses become increasingly important at higher redshifts. As the energy density of the CMB increases, the CMB photons will more efficiently scatter electrons, contributing more to cooling the plasma. The cooling in 4C 43.15 is still mainly through synchrotron losses, as $B_\mathrm{lobe} > B_\mathrm{CMB}$, but the difference is only a factor of $1.5$. This makes both synchrotron and IC losses important mechanisms for energy loss. The magnetic field strengths reported in \cite{Vaddi2019} increase with redshift and reach $\sim90\%$ of that found in 4C43.15, but also show a wide range of values. If the magnetic field strength of 4C 43.15 is not representative for the average magnetic field strength in HzRGs, then IC losses could be a significant contributor or start to dominate at higher redshifts or for sources with weaker magnetic fields in their lobes. Such a scenario, that increased IC losses
can explain the observed correlation, is also consistent with the findings of \cite{Morabito2018}.

\section{Conclusion}
The aim of this paper was to investigate whether increased IC losses, Malmquist bias, or environmental effects are the possible drivers behind the $\alpha-z$ relation. For the first time we demonstrate spatially resolved spectral modelling of a high-redshift object using low-frequency observations by combining high angular resolution ILT and VLA data from $55\ \mathrm{MHz}$ to $8.4\ \mathrm{GHz}$. We used the \textsc{brats} software package to fit a spectral ageing model to the data. A magnetic field strength of $B= 5.2\ \mathrm{nT}$ was estimated and best fitting injection indices of $\alpha^\mathrm{north}_\mathrm{inj} = -0.8$ and $\alpha^\mathrm{south}_\mathrm{inj} = -0.7$ were determined. Using these to fit a JP Tribble ageing model, we measure a young spectral age compared to local sources, with values of $\tau_\mathrm{spec} = 0.8 \pm 0.1\ \mathrm{Myr}$ and $\tau_\mathrm{spec} = 0.9 \pm 0.1\ \mathrm{Myr}$ for the northern and southern lobes.

Although no strong conclusions about the origins or the $\alpha-z$ correlation can be drawn from this source alone, our data supports the possibility that the increasing IC losses at higher redshifts is one of the drivers behind it. The strong magnetic field of 4C 43.15  seems plausible. In this case synchrotron losses still dominate over IC losses, but a magnetic field that is half the strength would already reverse this. Assuming that the jet power, radio luminosity and   magnetic field strength all scale with each other, the general high-redshift population may well be dominated by IC losses, gradually switching dominant mechanisms as redshift increases. Given the high luminosity of 4C 43.15 and the lack of data on a larger sample, we cannot rule out that the $\alpha-z$ relation is partially driven by Malmquist bias. Finally, we have ruled out that an increased density of the environment steepens the spectrum.

A handful of studies have now demonstrated LOFAR's unique capabilities for high-resolution imaging. In the future, high-resolution imaging will become routine. When combined with LOFAR's wide field of view, a large sample of spatially resolved high-redshift objects   will become available. This will allow us to extend analyses such as the one in this paper to a statistically significant sample and investigate the low-frequency structure and spectra of the general high-redshift population. Finally, the Low Band Antennas of the ILT will further push the boundaries, providing subarcsecond resolution at frequencies well below $100\ \mathrm{MHz}$.
\begin{acknowledgements}
We thank Sean Mooney, Rafaella Morganti and Raymond Oonk for their valuable input on early drafts of this manuscript. This paper is based (in part) on data obtained with the International LOFAR Telescope (ILT) under project code LT5\_006. LOFAR (van Haarlem et al. 2013) is the Low Frequency Array designed and constructed by ASTRON. It has observing, data processing, and data storage facilities in several countries, that are owned by various parties (each with their own funding sources), and that are collectively operated by the ILT foundation under a joint scientific policy. The ILT resources have benefited from the following recent major funding sources: CNRS-INSU, Observatoire de Paris and Université d'Orléans, France; BMBF, MIWF-NRW, MPG, Germany; Science Foundation Ireland (SFI), Department of Business, Enterprise and Innovation (DBEI), Ireland; NWO, The Netherlands; The Science and Technology Facilities Council, UK; Ministry of Science and Higher Education, Poland.
RJvW acknowledges support from the ERC Starting Grant ClusterWeb 804208.
This work has made use of the Dutch national e-infrastructure with the support of SURF Cooperative through grant e-infra 180169. JM acknowledges financial support from the State Agency for Research of the Spanish MCIU through the ``Center of Excellence Severo Ochoa'' award to the Instituto de Astrof\'isica de Andaluc\'ia (SEV-2017-0709) and from the grant  RTI2018-096228-B-C31 (MICIU/FEDER, EU)
\end{acknowledgements}

\bibliographystyle{aa}
\bibliography{bibliography}

\begin{appendix} 
\section{LOFAR LBA 55 MHz map}
\begin{figure}[h]
        \includegraphics[width=\columnwidth]{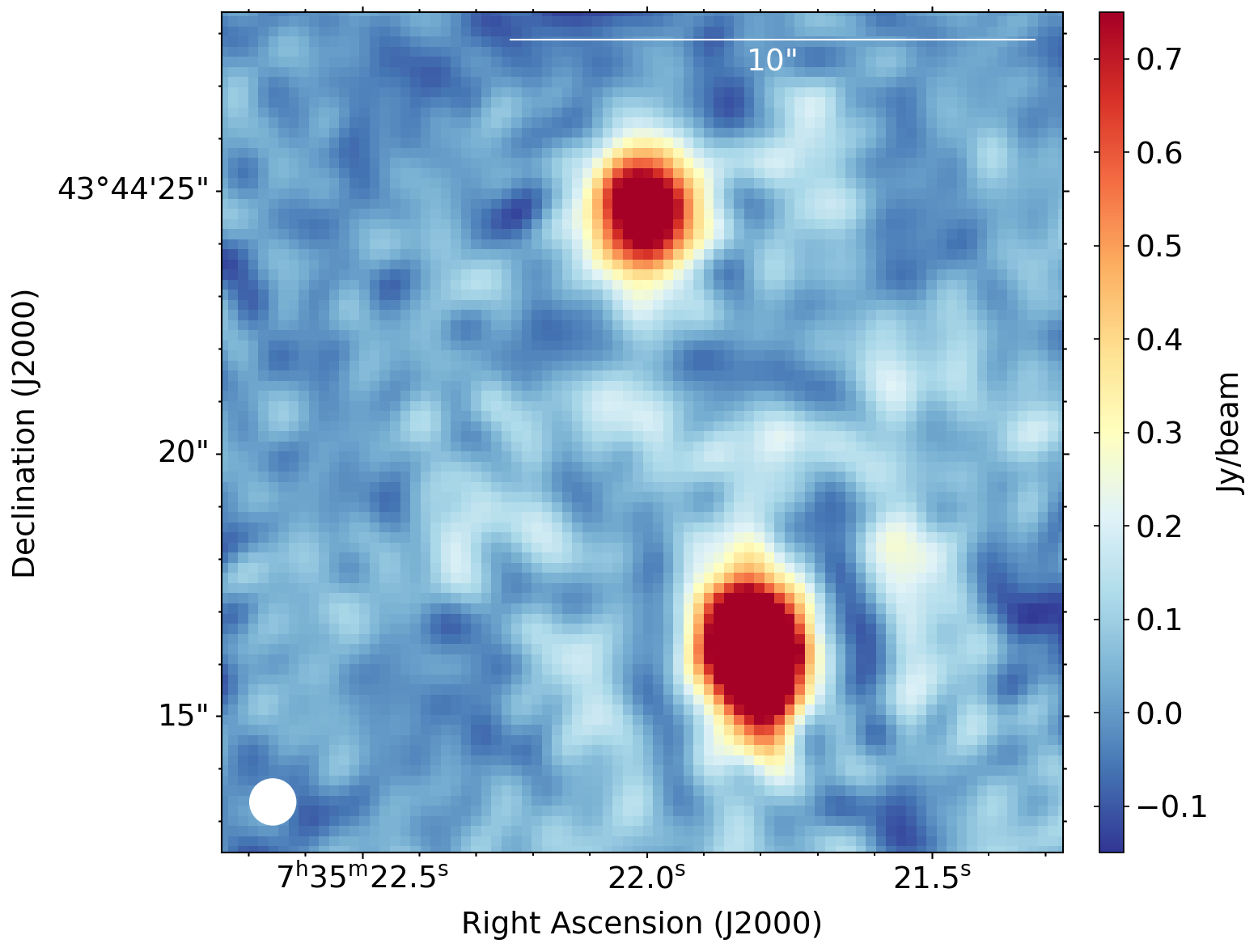}
        \caption{LOFAR LBA map at $55\ \mathrm{MHz}$ from \citet{Morabito2016}. The colour scale is shown on a linear stretch and ranges from $-0.15$ to $0.75\ \mathrm{Jy beam}^{-1}$. In the bottom left corner a white circle illustrates the restoring beam of $0.9''$ to which the image was smoothed. North is up and east is to the left.}
        \label{fig:lbamap}
\end{figure}

\section{Injection index and model fits}
\begin{figure}[h]
        \includegraphics[width=\columnwidth]{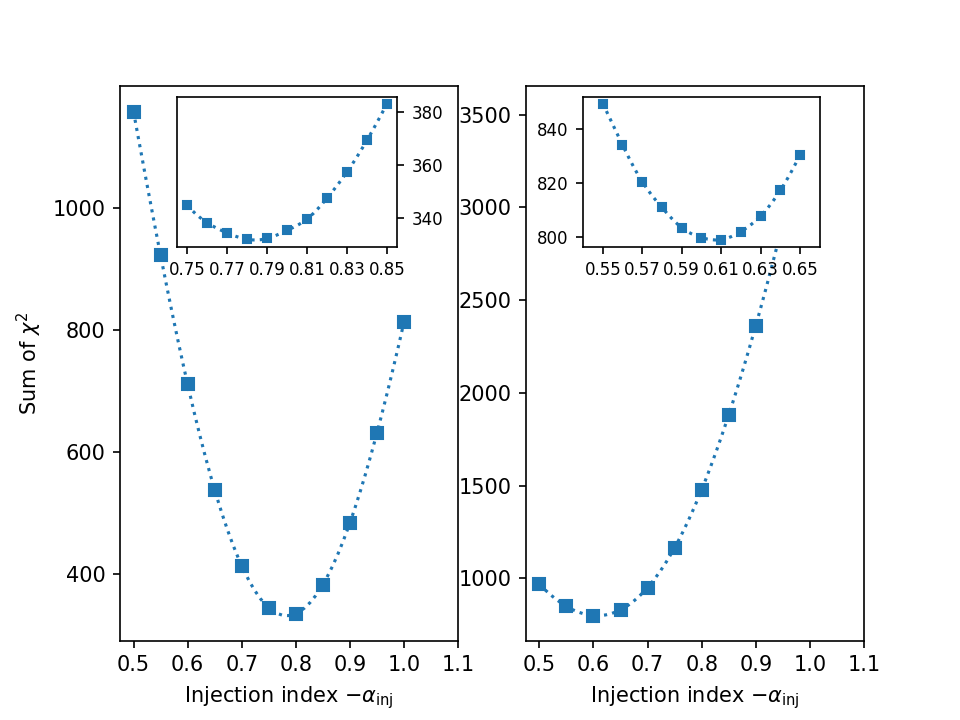}
        \caption{$\chi^2$ landscape of the initial broad search for the injection index $\alpha_\mathrm{inj}$. \textit{Left}: Northern lobe. \textit{Right}: Southern lobe. Squares indicate the explored values. Dotted lines are cubic spline fits for illustrative purposes. The large panels show the $\chi^2$ values for the broad search from $0.5$ to $1.0$. The  insets show the $\chi^2$ values for the detailed search in a narrower range of $0.75-0.85$ (left) and $0.55-0.65$ (right).}
        \label{fig:injectindex}
\end{figure}

\begin{figure*}
        \includegraphics[angle=-90,origin=c,width=\columnwidth]{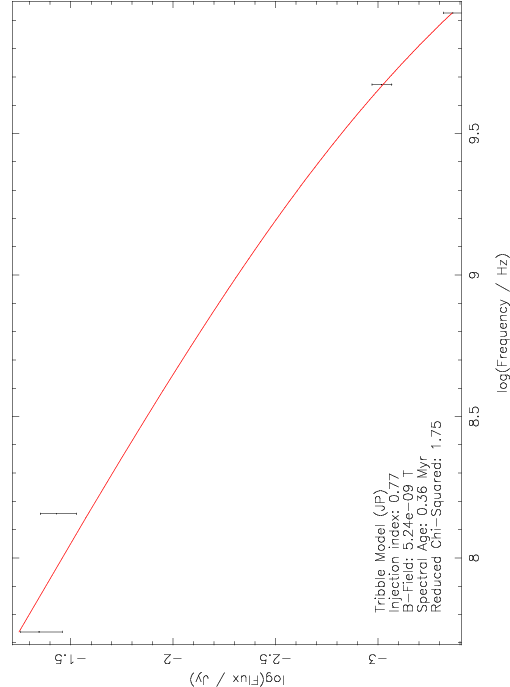}
        \includegraphics[angle=-90,origin=c,width=\columnwidth]{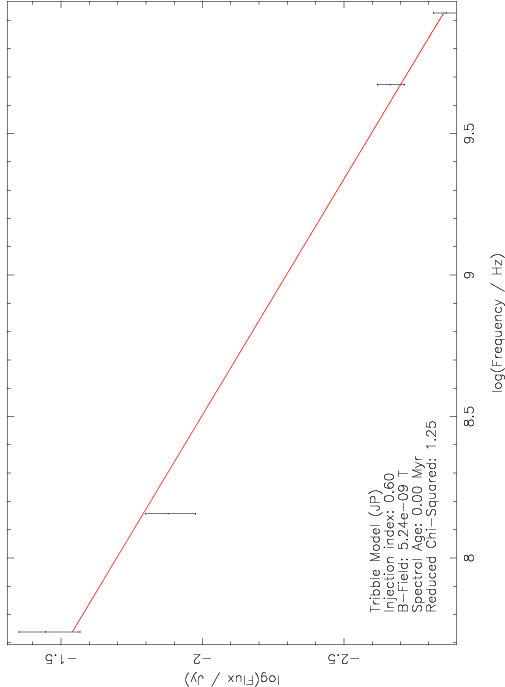}
        \includegraphics[angle=-90,origin=c,width=\columnwidth]{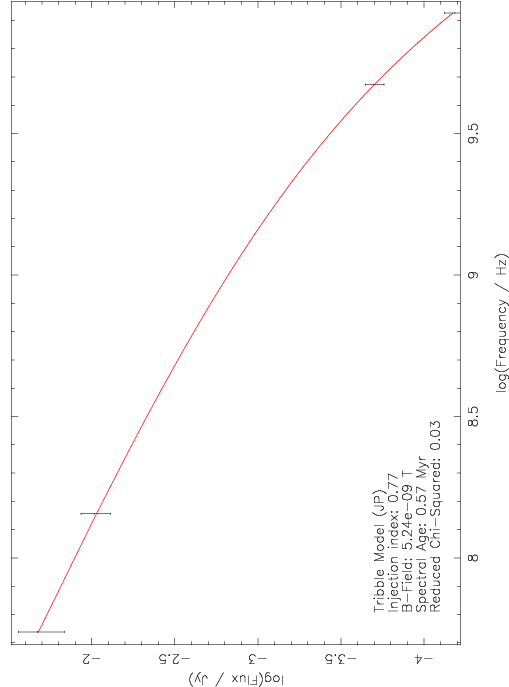}
        \includegraphics[angle=-90,origin=c,width=\columnwidth]{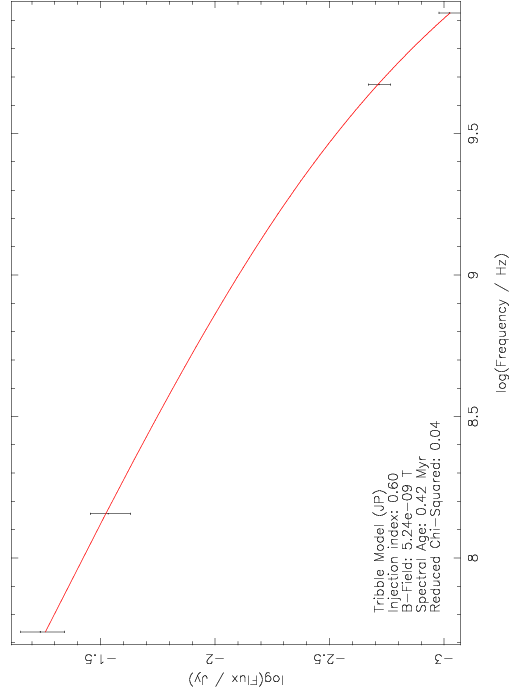}
        \includegraphics[angle=-90,origin=c,width=\columnwidth]{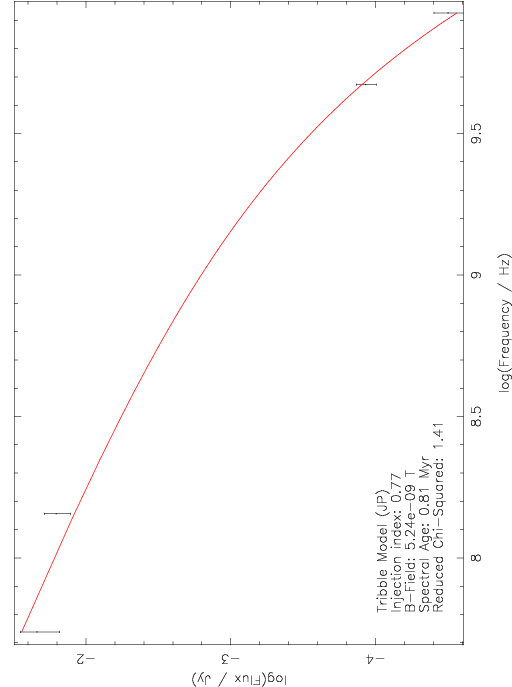}
        \includegraphics[angle=-90,origin=c,width=\columnwidth]{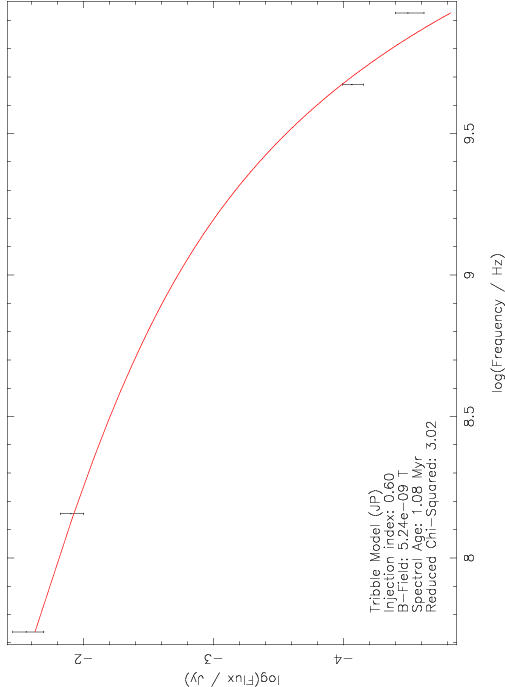}
        \caption{Model fits of the JP Tribble spectral ageing model for the northern (\textit{left}) and southern (\textit{right}) lobes. Fits in a region with the youngest spectral age (\textit{top}), lowest reduced $\chi^2$ (\textit{middle}), and oldest spectral age (\textit{bottom}) are shown. The black points with error bars indicate the flux density data points in that region and the red line shows the model fit. In the lower left corner of each panel the fitted model, best fitting injection index, magnetic field strength and spectral age in that region, and the reduced $\chi^2$ values are listed.}
        \label{fig:modelfits}
\end{figure*}
\end{appendix}
\end{document}